\documentclass[aps,prb,twocolumn,showpacs,superscriptaddress,floatfix]{revtex4-1}
\usepackage{epsfig}
\usepackage{graphicx}
\usepackage{amsfonts}
\usepackage[figuresright]{rotating}
\usepackage{amssymb}
\usepackage{amsmath,amsthm}

\def\be{\begin{equation}} \def\ee{\end{equation}}
\def\bea{\begin{eqnarray}} \def\eea{\end{eqnarray}}

\def\nn{\nonumber}

\begin{document}
\title{Vortex Lattices in the Superconducting Phases of Doped Topological Insulators and Heterostructures}
\author{Hsiang-Hsuan Hung}
\affiliation{Department of Electrical and Computer Engineering,
University of Illinois, Urbana, Illinois 61801}
\affiliation{Micro and Nanotechnology Laboratory, University of Illinois, 208 N. Wright St, Urbana IL 61801}
\author{Pouyan Ghaemi}
\affiliation{Department of Physics, University of Illinois, Urbana, Illinois 61801}
\author{Taylor L. Hughes}
\affiliation{Department of Physics, University of Illinois, Urbana, Illinois 61801}
\author{Matthew J. Gilbert}
\affiliation{Department of Electrical and Computer Engineering,
University of Illinois, Urbana, Illinois 61801}
\affiliation{Micro and Nanotechnology Laboratory, University of Illinois, 208 N. Wright St, Urbana IL 61801}

\begin{abstract}
Majorana fermions are predicted to play a crucial role in condensed matter realizations of topological quantum computation. These heretofore undiscovered quasiparticles have been predicted to exist at the cores of vortex excitations in topological superconductors and in heterostructures of superconductors and materials with strong spin-orbit coupling. In this work we examine topological insulators with bulk $s$-wave superconductivity in the presence of a vortex-lattice  generated by a perpendicular magnetic field. Using self-consistent Bogoliubov-de Gennes, calculations we confirm that beyond the semi-classical, weak-pairing limit that the Majorana vortex states appear as the chemical potential is tuned from either side of the band edge so long as the density of states is sufficient for superconductivity to form. Further, we demonstrate that the previously predicted vortex phase transition survives beyond the semi-classical limit. At chemical potential values smaller than the critical chemical potential, the vortex lattice modes hybridize within the top and bottom surfaces giving rise to a dispersive low-energy mid-gap band. As the chemical potential is increased, the Majorana states become more localized within a single surface but spread into the bulk toward the opposite surface. Eventually, when the chemical potential is sufficiently high in the bulk bands, the Majorana modes can tunnel between surfaces and eventually a critical point is reached at which modes on opposite surfaces can freely tunnel and annihilate leading to the topological phase transition previously studied in the work of  Hosur \emph{et al.}\cite{hosur2011}.
\end{abstract}
\maketitle


\section{Introduction}
\label{sect:intro}

Majorana fermions, quasi-particle excitations which are their own antiparticle, were originally proposed in high-energy physics but \cite{majorana1937} have now arrived at the forefront of condensed matter physics where they serve as non-Abelian anyons which form the backbone of topological quantum computing architectures\cite{ivanov2001,kitaev2001,wilczek2009,hughes2011,dassarma2006,read2000,kitaev2001,nayak2008}. Within condensed matter physics, there exist many candidate systems which are predicted to harbor Majorana fermions. One of the earliest of such candidates is the fractional quantum Hall effect at filling factor $\nu=\frac{5}{2}$ \cite{willett1987} the physics of which may be described by the Moore-Read pfaffian wavefunction\cite{moore1991}. While this state is yet to be experimentally confirmed, tantalizing evidence observed in tunneling in quantum constrictions points to the fact that the $\nu=\frac{5}{2}$ fractional quantum Hall state does possess non-Abelian statistics as would be necessitated by the presence of Majorana fermions\cite{radu2008}. Beyond the fractional quantum Hall states, other possible systems thought to contain Majorana fermions are the $p_{x}+ip_{y}$ superconductors\cite{moore1991,read2000,ivanov2001,kitaev2003,cheng2009} where the relevant Majorana modes are predicted to appear as bound-states on exotic half-quantum vortices, which were recently observed in magnetic force microscopy experiments performed in Sr$_2$RuO$_4$\cite{jang2011}. In addition to fractional quantum Hall states and superconductors, there has been an abundance of proposals to realize Majorana fermions in materials with strong-spin orbit coupling. Notable examples are proximity induced superconductivity in 3D topological insulators (TIs)\cite{fu2008}, bulk superconductivity in doped TIs \cite{hosur2011} and semiconductors coupled proximity coupled to s-wave superconductors\cite{sau2010,alicea2010,lutchyn2010,oreg2010}. Indeed, the latter proposals have led to exciting measurements in high mobility quantum wire - $s$-wave superconductor systems\cite{mourik2012}. 

In this article we will focus on the two mechanisms proposed in TI materials. As is now well-known, TIs are materials which possess an insulating bulk but contain robust metallic states that are localized on their surfaces\cite{kane2005,kane2005sf,bernevig2006, konig2007,fu2007prl,moore2007,roy2009,kane2010,hsieh2008,chengyl2009,hsieh2009}. We will consider time-reversal invariant 3D topological insulators which harbor an odd number of massless Dirac cones on each surface. 
As mentioned, there currently exist two proposals that utilize topological insulators as a platform for the observation of Majorana fermions.
The first of which is to consider an $s$-wave superconductor/topological insulator heterostructure in which  a superconductor is coupled to the topological insulator via the proximity effect and subjected to a vortex-producing magnetic field.\cite{fu2008} In Fu and Kane's pioneering work,\cite{fu2008} they show that in the $s$-wave superconductor/TI heterostructure, the interface between the TI and the superconductor behaves similar to a spinless chiral $p$-wave superconductor yet without breaking time reversal symmetry. As such, Majorana fermions will reside in the vortex cores\cite{volovik1999,fu2008} so long as the quantized magnetic flux lines penetrating the system are broad\cite{chiu2011}. 

On the other hand, another strategy to realize Majorana fermions
is to consider vortex bound-states  in a 3D TI with bulk
$s$-wave superconductivity\cite{hosur2011,hughes2011,qi2010}. Bulk
superconductivity in doped topological insulators has been observed
in recent experiments that dope Bi$_2$Se$_3$ with copper.\cite{hor2010,hor2011,wray2011}
For this particular material the nature of the order parameter is still under debate, but the two most probable options are $s$-wave, or an inter-orbital topological pairing parameter\cite{fu2010}. There is an opportunity to observe Majorana fermions in both cases, but for the purpose of this work we will only consider the $s$-wave case. 
 Recent
work\cite{hosur2011} reveals that, while doped topological
insulators that develop $s$-wave pairing may harbor Majorana bound states in the vortices,
 the Majorana fermions do not
survive for all doping levels.  Specifically, there exists a
critical chemical potential, $\mu_c$, at which point the system
undergoes a topological (vortex) phase transition. This  phase transition can
be regarded as a topology change in the 1D electronic structure of
vortex lines from a system which supports gapless end states to one
that does not. Therefore, only at chemical potentials below the
critical value, the doped superconducting Bi$_2$Se$_3$ supports
Majorana modes at the vortex ends (places where vortex lines intersect the surface). The work of Ref. \onlinecite{hosur2011} provided a semiclassical treatment in the infinitesimal pairing limit, but such an approach
might be inadequate to capture important quantum effects. One such effect that cannot be determined this way is the zero-point energy contribution of the vortex core states which
could shift the states away from the zero energy gapless point and invalidate the previous analysis.
 In the weak-pairing limit the physics is determined by the structure exactly at the Fermi-surface, and it is possible that the energy spectrum away from the Fermi-level can serve to renormalize the location of the critical point. If the critical point is sufficiently shifted toward the band-edge, it could be that there is never a viable doping-range over which the Majorana fermions can be observed. Calculations provided in this article go beyond the semi-classical, and infinitesimal weak-pairing limit are
thus essential to confirm the previous results. 

In both TI based approaches we have thus far discussed, there is an additional assumption underlying the resultant physical predictions,
namely that the vortices are completely isolated.  However, this may not be the most
appropriate or experimentally relevant picture for the observation
of Majorana states in either type of TI approach. Thus, in this article, we examine the behavior of 3D topological insulators with
bulk $s$-wave superconductivity in the vortex lattice limit as a
function of doping level. The paper is organized in the following
fashion: In Sections \ref{sect:hamiltonian}, and \ref{sect:bulkSC}, we introduce the 3D
topological insulator model Hamiltonian which is used for each of
the subsequent calculations, and the background for the self-consistent calculations respectively. In Section \ref{sect:vortexlattice}, we
present the results of our calculations for three separate geometries (a)periodic boundary conditions with vortex rings (b) open boundary conditions with vortex lines terminating on the TI surface (c) an inhomogeneously doped heterostructure with open boundary conditions. We find that as the
chemical potential is adiabatically moved from the gap into the
bulk bands, the Majorana states form when the density of states reaches 
is large enough to support a well-formed superconducting gap.  As the chemical potential moves past
this onset value we find that the vortices are localized on the surfaces but hybridize with neighboring vortices on the same surface giving rise to a dispersive low-energy quasi-particle spectrum.
As the
chemical is pushed further into the bands, we
find a critical chemical, $\mu_{t}$, at which inter-surface tunneling is enabled through a gapless channel on the vortex line. The value is renormalized from that stated in Ref. \onlinecite{hosur2011}, but we find that even for strong attractive interactions that $\mu_t$ remains at a finite value of the bulk-doping. After the chemical potential
exceeds $\mu_{t}$, we find that  a gap opens in the
spectrum and there are no longer any low-energy localized modes remaining. Additionally, in Section \ref{sec:finiteT}, we evaluate
the superconducting gap equation in order to determine the relevant
temperature scale on which these effects may be observed. Finally,
in Section \ref{sect:conclusion} we summarize our findings
and conclude.

\section{Model Hamiltonian of a 3D Topological Insulator}
\label{sect:hamiltonian}


We will use a minimal, four-band Dirac-type model which, with the proper choice of parameter values, captures the bulk, low-energy physics of known TI materials such as Bi$_2$Se$_3$\cite{zhang2009,liu2010}:
\bea \label{eq:ham0}
H&=&\sum_{\vec{r}}\Big\{
\Psi^{\dag}_{\vec{r}}H_m\Psi_{\vec{r}}+\sum_{\vec{\delta}}\Psi^{\dag}_{\vec{r}}H_{\vec{\delta}}\Psi_{\vec{r}+\vec{\delta}}
\Big\},\eea  \bea H_m=\mathbb{M} \Gamma^0, \ \
H_{\vec{\delta}}=\sum_{\vec{\delta}} \frac{b\Gamma^0+i \gamma \vec{\delta}\cdot \vec{\Gamma}}{2a^2},
\eea
\noindent where $\Psi_{\vec{r}}=(c_{A,\uparrow,\vec{r}} \
c_{A,\downarrow,\vec{r}} \ c_{B,\uparrow,\vec{r}} \ c_{B,
\downarrow,\vec{r}})^T$ is a four-component spinor with $A/B$ and $\uparrow/\downarrow$ labeling orbital and physical spin respectively so that $c^\dagger_{\alpha,\sigma,\vec{r}}$ is the creation operator for an electron with spin $\sigma$ in orbital $\alpha$ at position $\vec{r},$  $\vec{\delta}= \pm a \hat{x}, \ \pm a \hat{y}, \
\pm a \hat{z}$  are vectors that connect nearest neighbors on a simple cubic lattice with lattice constant $a,$ the vector $\vec{\Gamma}=\Gamma^x\hat{x}+\Gamma^y\hat{y}+\Gamma^z\hat{z}$ with
$\Gamma^{\alpha}=\tau^{x} \otimes \sigma^{\alpha}$ and
$\Gamma^0=\tau^z \otimes \mathbb{I}$, where $\alpha=x,\ y,\ z$;
$\tau^{\alpha}$ and $\sigma^{\alpha}$ are $2 \times 2$ Pauli
matrices acting on orbital and spin degrees of freedom,
respectively. We also define $\mathbb{I}$ as the $2 \times 2$ identity matrix and
 $\mathbb{M}=m-3b/a^2$  as the mass parameter which controls the magnitude of the bulk band gap. The TI/trivial
insulator phase depends on the chosen values for the parameters $m$ and $b$ and the TI phase has $m/b>0$ while the trivial phase has $m/b<0$. By tuning the material parameters $\gamma, b, m,$ and $a$ in Eq.  (\ref{eq:ham0}) one can model the low energy effective model for the common binary TI materials\cite{zhang2009,liu2010}. Since we only address the qualitative effects stemming from the TI phase we will fix the parameters to be $b=a^2(1eV),$ $\gamma=a(1 eV)$ and  $m=1.5$ eV in terms of the lattice constant $a$ so that $\mathbb{M}=-1.5$ eV thereby ensuring that we are in the TI phase.

With translation invariance and periodic boundary conditions in all $x$, $y$ and $z$ directions, it is often more convenient to work in momentum space. In this case, we expect to see no gapless states due to the lack of a boundary. The Fourier transformed Dirac Hamiltonian in momentum space
may be written as  \bea  \label{eq:ham0k} H= \sum_{\vec{k}}
\Psi^{\dag}_{\vec{k}} H_0(\vec{k}) \Psi_{\vec{k}},\eea
where
$H_0(\vec{k})$ in Eq. (\ref{eq:ham0k}) is a $4 \times 4$ matrix which we may write as
\begin{widetext}
\bea
\label{eq:hamk} H_0(\vec{k})
&=&\left (
\begin{array} {c c c c}
\mathbb{M}+g(\vec{k}) & 0 & \sin{k_z} &\sin{k_x}-i\sin{k_y} \\
0 & \mathbb{M}+g(\vec{k}) & \sin{k_x}+i\sin{k_y} &-\sin{k_z} \\
\sin{k_z} & \sin{k_x}-i\sin{k_y} & -[\mathbb{M}+g(\vec{k})] &0 \\
\sin{k_x}+i\sin{k_y} & -\sin{k_z} & 0 & -[\mathbb{M}+g(\vec{k})]
\end{array}  \right ),
 \eea
 \end{widetext}
with $g(\vec{k})=\cos{k_x}+\cos{k_y}+\cos{k_z}$.
If we instead choose open boundaries along one direction there will be robust gapless edge states on those boundary surfaces for the same choice of model parameters.


\section{Topological Insulators with Bulk $S$-Wave Superconductivity}
\label{sect:bulkSC}
In this work, we are interested in the properties of doped TIs which become intrinsically superconducting at low temperature.  In a doped topological insulator, like any other metal, when the chemical potential is in the conduction or valence band an attractive interaction will lead to the formation of superconductivity and generate a superconducting gap at the Fermi surface.
In order to study the formation of superconductivity in doped TI we add an attractive Hubbard-type density-density interaction to the Hamiltonian in Eq.  (\ref{eq:ham0}) :
\bea H_{int}=-|U|\sum_{\vec{r}}n_{\uparrow,\vec{r}}n_{\downarrow,\vec{r}}
\eea
where $n_{\sigma,\vec{r}}=c^{\dag}_{A,\sigma,\vec{r}}c^{\phantom{\dagger}}_{A,\sigma,\vec{r}}+c^{\dag}_{B,\sigma,\vec{r}}c^{\phantom{\dagger}}_{B,\sigma,\vec{r}}$ and the parameter $-|U|$ represents the attractive intra-orbital interaction.

At the mean-field level the interaction term may be decoupled as\cite{degennes1966}:
\bea -|U| \sum_{\alpha,\vec{r}} \Big\{ \Delta_{\alpha,\vec{r}}^*
c_{\alpha,\downarrow,\vec{r}}c_{\alpha,\uparrow,\vec{r}}+
\Delta_{\alpha,\vec{r}}
c^{\dag}_{\alpha,\uparrow,\vec{r}}c^{\dag}_{\alpha,\downarrow,\vec{r}}
  - |\Delta_{\alpha,\vec{r}}|^2 \Big\},\nn  \eea
where
$\Delta_{\alpha,\vec{r}}=\langle c_{\alpha,\downarrow,\vec{r}}
c_{\alpha,\uparrow,\vec{r}}\rangle$ is the standard intra-orbital s-wave pairing order
parameter.
Combining this with Eq. (\ref{eq:ham0}), we get the Bogoliubov-de Gennes (BdG) Hamiltonian:
\bea \label{eq:bdG} &H_{BdG}& = \sum_{\vec{r}}
\Phi^{\dag}_{\vec{r}} \left (
\begin{array} {c c}
H_m-\mu_{\vec{r}} & \Delta(\vec{r})   \\
\Delta(\vec{r})^{\dag}  & -H^*_m+\mu_{\vec{r}}
\end{array}  \right ) \Phi_{\vec{r}} \nn \\
&+& \sum_{\vec{r},\vec{\delta}} \Phi^{\dag}_{\vec{r}} \left (
\begin{array} {c c}
H_{\vec{\delta}} & 0  \\
0 & -H^*_{\vec{\delta}}
\end{array}  \right ) \Phi_{\vec{r}+\vec{\delta}},
 \eea
where $\Phi_{\vec{r}}=(\Psi_{\vec{r}}\ , \Psi^{\dag}_{\vec{r}})^T$ is now an 8-component Nambu spinor, and $\Delta(\vec{r})$ denotes a $4\times 4$ pairing matrix. In this expression, the interaction $-|U|$ has been absorbed into the pairing matrix
$\Delta(\vec{r})$, which we write as
\bea \label{eq:intraorbitalmatrix}
\Delta(\vec{r})&=&-|U|\left
(
\begin{array} {c c c c}
0&\Delta_{A,\vec{r}}&0&0  \\
-\Delta_{A,\vec{r}}&0&0&0 \\
0&0& 0& \Delta_{B,\vec{r}}\\
0&0&-\Delta_{B,\vec{r}}&0
\end{array}  \right )
\eea
To study the bulk superconductivity we will assume $\mu_{\vec{r}}=\mu$ is uniform throughout the material for simplicity.

The BdG Hamiltonian of Eq.
 (\ref{eq:bdG}) can be diagonalized by applying a Bogoliubov transformation as\cite{degennes1966}
\bea \label{eq:bdgeq}
{\Psi_{\vec{r}} \choose \Psi^{\dag}_{\vec{r}} }= \sum_{n} \left (
\begin{array} {cc}
u_{n,\vec{r}} & -v^{*}_{n,\vec{r}} \\
v_{n,\vec{r}} & u^{*}_{n,\vec{r}}
\end{array}  \right )
{\gamma_{n} \choose \gamma^{\dag}_{n}}, \eea where $n$ labels the
eigenstate index. Plugging the transformation into Eq.
(\ref{eq:bdG}), we have \bea \label{eq:diagonalization} &H_{BdG}& \sum_n \left (
\begin{array} {cc}
u_{n,\vec{r}} & -v^{*}_{n,\vec{r}} \\
v_{n,\vec{r}} & u^{*}_{n,\vec{r}}
\end{array}  \right )\nn \\ &=&\sum_n \left (
\begin{array} {cc}
E_n & 0 \\
0 & -E_n
\end{array}  \right )
\left (
\begin{array} {cc}
u_{n,\vec{r}} & -v^{*}_{n,\vec{r}} \\
v_{n,\vec{r}} & u^{*}_{n,\vec{r}}
\end{array}  \right ).  \eea
This indicates that the eigenvectors associated with $E_n$ ($-E_n$)
of the above BdG equations are $(u_{n,\vec{r}},v_{n,\vec{r}})^T$
[$(-v^*_{n,\vec{r}},u^*_{n,\vec{r}})^T$]. The mean-field pairing order
parameters are obtained via  \bea
\Delta_{\alpha,\vec{r}}&=&\langle c_{\alpha,\downarrow,\vec{r}}
c_{\alpha,\uparrow,\vec{r}} \rangle \nn \\ &=&\sum_n u_{n,\vec{r}}
v^*_{n,\vec{r}} \tanh{\frac{\beta E_n}{2}}, \eea where
$\beta=1/k_BT$. Once the pairing order parameter are determined initially
it is plugged back into the BdG Hamiltonian given in Eq. (\ref{eq:bdG}) and then $H_{BdG}$ is diagonalized again as shown in Eq. (\ref{eq:diagonalization}). The process continues until we reach self-consistency and
we have a convergent $\Delta_{\alpha,\vec{r}}$ for all $\vec{r}$. We note that in our numerical calculations we use a small, non-zero temperature in order to avoid divergences but this temperature is much smaller than the superconducting gap so as not to affect the physical results.

\begin{figure}[t]
\epsfig{file=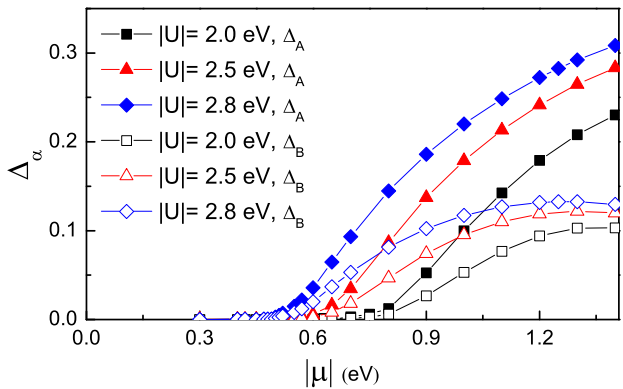,clip=0.7,width=0.95\linewidth,angle=0}
\caption{(color online).  Intra-orbital pairing order parameters $\Delta_{A}$ (solid symbols) and $\Delta_B$ (hollow symbols) vs $|\mu|$ at different $|U|$. $|\mu|$ and $|U|$ are in units of eV. The mass term $\mathbb{M}=-1.5$ eV.
The system contains periodic boundary conditions in $x$, $y$ and $z$ directions. The simulations are performed on a lattice grid of size $80a\times 80a\times 10a$.}
\label{fig:periodiczTIdifferentU}
\end{figure}

In Fig. \ref{fig:periodiczTIdifferentU} we show the
self-consistently determined intra-orbital pairing order parameter in
the bulk as a function of $|\mu|$ at different $|U|$, where, due to
translation invariance, $\Delta_{\alpha,\vec{r}}=\Delta_{\alpha}.$
In this paper we will only consider p-doping ($\mu<0$), but the
particle-hole symmetry of the model Hamiltonian in Eq.
(\ref{eq:ham0}) makes the electron-doped case similar in nature.
With $\mathbb{M}=-1.5$ eV in Eq. (\ref{eq:ham0}), the top of the
bulk valence band is located at $\mu_v=-0.5$ eV and the total size
of the insulating gap is $1.0$ eV. We see from Fig.
\ref{fig:periodiczTIdifferentU} that when the chemical potential is
in the gap where there is no carrier density with which to form
Cooper pairs and the resulting pairing potential is zero. When the
chemical potential enters the valence band a Fermi-surface develops,
and low-energy states become available to pair. However, when the
density of states at the Fermi-level is insufficient, the size of
the pairing potential will continue to be exponentially small. As we
see in Fig.  \ref{fig:periodiczTIdifferentU}, a significant pairing
potential does not form until $|\mu|$  is well above the valence
band edge, $|\mu_v|.$

This result matches standard BCS phenomenology and represents the point of inception for the remainder of the paper. To be specific, in Ref. \onlinecite{hosur2011} Hosur et al. used a semi-classical treatment to show that a vortex in the superconducting phase of a doped topological insulator exhibits a topological phase transition as the chemical potential is tuned through a critical value.   The two phases separated by this transition are gapped and differ by the presence or absence of Majorana modes at the ends of the vortex \emph{i.e.} where the vortex line intersects the surface of the TI. At the transition point the vortex line becomes gapless and provides a channel which allows the Majorana modes to annihilate one another by tunneling in-between the opposing surfaces. In their treatment, however, there is an assumption of adiabaticity as it is always assumed that at any chemical potential other than the critical chemical potential, there is no gapless channel to hybridize the Majorana mode. This seems innocuous, but one has to remember that the arguments rely on the adiabatic connection between a gapped insulating phase and a gapped superconducting phase. The assumption enters when one considers the behavior of the system as the chemical leaves the insulating gap and enters the bulk bands. Although this is a reasonable assumption within which to theoretically study the vortex phase transition, one may then ask what happens in the region where the chemical potential is not large enough to form a significant pairing potential, and there is finite density of gapless modes in the bulk. In other words, how does the Majorana mode emerge out of the bulk gapless states? This question is certainly relevant for experiments where a finite size TI sample  is used. Our self-consistent solution of the BdG equations in the vortex lattice can present a clearer picture of the appearance of the Majorana modes and the vortex phase transition than the previous semi-classical analysis. 

\section{Vortex Lattices in Superconducting Phase of Doped Topological Insulators}
\label{sect:vortexlattice}

The self-consistent BdG formalism is in a real space basis and thus
can be also used to study the vortices in the superconducting phase
where the order parameter will be non-uniform. To induce vortices,
we consider the system under a uniform magnetic field
$\vec{B}=B\hat{z}.$ When electrons are hopping on the $xy$-plane
this generates a  Peierls phase factor,  and the BdG Hamiltonian
becomes\cite{vafek2001,chiu2011} \bea \label{eq:newbdg} &H_{BdG}& =
\sum_{\vec{r}} \Phi^{\dag}_{\vec{r}} \left (
\begin{array} {c c}
H_{m}-\mu_{\vec{r}} & \Delta(\vec{r})   \\
\Delta^{\dag}(\vec{r}) & -H^*_{m}+\mu_{\vec{r}}
\end{array}  \right ) \Phi_{\vec{r}}  \nn \\
&+& \sum_{\vec{r},\delta} \Phi^{\dag}_{\vec{r}} \left (
\begin{array} {c c}
H_{\delta}e^{-i\eta_{\vec{r}}} & 0  \\
0 & -H^*_{\delta}e^{i\eta_{\vec{r}}}
\end{array}  \right ) \Phi_{\vec{r}+\delta}\nn \\ \eea
\noindent where $\eta_{\vec{r}}$ denotes the extra
phase given by the vector potential $\vec{A}({\vec{r}})$ induced by
the magnetic field through $\vec{B}=\nabla \times A({\vec{r}})$:
\bea
\label{eq:vectorpotential}\eta_{\vec{r}}=\frac{\pi}{\Phi_0}\int^{\vec{r}+\delta}_{\vec{r}}\vec{A}_{\vec{r}'}\cdot d \vec{r}',
\eea
\noindent where $\Phi_0$ is the superconducting flux quantum; $\Phi_0=\frac{h}{2e}$.
In the following discussion, we choose the Landau gauge, i.e.
$\vec{A}(\vec{r})=(A_x,A_y)=(0,Bx)$.

We will treat the system as a type-II superconductor in a vortex lattice state. In each magnetic unit cell, the amount of magnetic
flux is $2\Phi_0$, so that each unit cell carries two superconducting vortices\cite{wang1995}.  We designate the size of each magnetic unit cell as
$l_xa \times l_ya \times l_za$ using the integers $l_i$ to denote the number of lattice sites in each spatial direction. For our choice of geometry we will use square vortex lattices, and fix $l_x=l_y/2$.  The
corresponding magnetic field magnitude is  \bea
B=\frac{2\Phi_0}{l_x l_y a^2}.\eea
From this relation we can observe that stronger magnetic
fields bring smaller magnetic unit cells, in which vortices are
closer each other. Therefore, the dilute vortex limit comes from applying
very weak magnetic fields. As is standard for lattice calculations with uniform field, in order to see experimentally reasonable field sizes we would need to use a very large number of unit cells as, for example, the case when $l_x=l_y=1$ gives a magnetic field on the order of thousands of Tesla. For our system sizes we have an un-physically large magnetic field on the order of $10^3$ T assuming a lattice constant of $1\AA.$ This, however, will not affect the qualitative physics in which we are interested and we will not worry about this issue any further.

\begin{figure}[t]
\epsfig{file=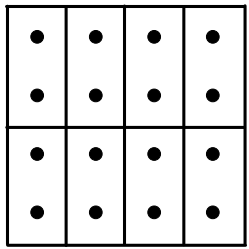,clip=0.7,width=0.38\linewidth,angle=0}
\caption{A $4 \times 4$  vortex lattice. In this example,
the number of magnetic unit cells is $N_x \times N_y =4 \times 2$.
Each black solid circle denotes a vortex location and each magnetic
unit cell contains two vortices.}
\label{fig:vortexlattice}
\end{figure}

We choose the entire system size as $L_xa \times L_ya \times l_za$
such that there are $N_x \times N_y$ magnetic unit cells, where
$N_{x}=L_{x}/l_{x}$ and $N_{y}=L_{y}/l_{y}$ and $N_x=2N_y$. Since
each magnetic unit cell carries two vortices, the $L_xa \times L_ya
\times l_za$ vortex lattice contains $2N_xN_y$ vortices. In Fig.
\ref{fig:vortexlattice}, we show a schematic illustration of a
$4\times 4$ square vortex lattice. By tuning sizes of the magnetic
unit cells, we can study the vortex lattice at different external
magnetic fields. In this paper, we set $l_x > 8$ to avoid strong
overlap between vortices but  $l_x \le 12$  due to computational
limitations.

We consider a system with periodic boundary conditions along the $x$ and $y$ directions. Although the vortices break lattice translation invariance, we still have magnetic
periodic boundary conditions for vortex lattices.
In addition to the phases given by
vector potentials $e^{i\eta_{\vec{r}}}$, the magnetic periodic boundary conditions also contribute another phase factor when
electrons are hopping across unit cell boundaries\cite{zak1964,zak1964b}.
Suppose that in a 2D  vortex lattice, the translation
vector in units of $a$ is written as
$\vec{\textrm{R}}=(\textrm{X}l_x,\textrm{Y}l_y)$, where
$\textrm{X}=0,\cdots,N_x-1$ and $\textrm{Y}=0,\cdots,N_y-1$ are
integers. The coordinate of an arbitrary lattice site can be
expressed as $\vec{r}+\vec{\textrm{R}}$, where $\vec{r}=(x,y)$
denotes the coordinate in units of the lattice site, $a$, within a magnetic unit cell, i.e. $1\le x \le l_x$ and $1 \le y \le l_y$. Under the magnetic periodic boundary conditions, we can define the relation of the magnetic Bloch wave functions\cite{han2010} which have a periodic structure written as\cite{han2010,hung2012} \bea \label{eq:magneticbloch} \left
(
\begin{array} {c}
u_{n}(\vec{r}+l_x\hat{x})  \\
v_{n}(\vec{r}+l_x\hat{x})
\end{array}  \right ) &=&  e^{ik_x}\left (
\begin{array} {c}
e^{2\pi i \frac{y}{l_y}}u_{n}(\vec{r}) \\
e^{-2\pi i \frac{y}{l_y}}v_{n}(\vec{r})
\end{array}  \right ), \nn \\
\left (
\begin{array} {c}
u_{n}(\vec{r}+l_y\hat{y})  \\
v_{n}(\vec{r}+l_y\hat{y})
\end{array}  \right ) &=&  e^{ik_y}\left (
\begin{array} {c}
u_{n}(\vec{r}) \\
v_{n}(\vec{r})
\end{array}  \right ).
\eea Here $k_x=\frac{2\pi \textrm{X}}{N_x}$ and $k_y=\frac{2\pi
\textrm{Y}}{N_y}$ represent the $x$ and $y$ components of the magnetic Bloch wavevector. The phases $e^{ik_x}$ and $e^{ik_y}$
 arise from hopping to neighboring cells. Additionally,
$e^{\pm2\pi i \frac{y}{l_y}}$ is provided by the
magnetic periodic boundary conditions, or quasi-periodic
boundary conditions\cite{han2010}.   The BdG eigenstates $(u_{n,\vec{r}},v_{n,\vec{r}})^T$
satisfy magnetic translation invariance under Eq. (\ref{eq:magneticbloch}).

The on-site pairing potential can be
expressed as $\Delta_{\alpha,\vec{r}}=|\Delta_{\alpha,\vec{r}}| e^{i
\phi(\vec{r})}$ with a phase $e^{i \phi(\vec{r})}$ and amplitude $|\Delta_{\alpha, \vec{r}}|$.
In the presence of vortices, both the pairing potential and the phase are site-dependent.
The superconducting order parameters are suppressed near the vortex cores, and are restored to the bulk values
away from the vortex cores. The spatial form of the pairing order parameters $\Delta_{\alpha,\vec{r}}$ are determined
self-consistently.  We consider different attractive-Hubbard interaction strengths $|U|$ and uniform doping-levels $|\mu|$ distributed through the entire bulk.

We study two different geometries for the vortex lattice. First we consider vortices oriented in the z-direction (along the applied magnetic field) with periodic boundary conditions along the $x,y,z$ directions which yields vortex rings looping around the z-direction. In this geometry we study the vortex phase transition where the vortex modes become gapless along the vortex rings.  The second geometry we consider has open boundaries in the $z$-direction. In this case the vortex lines terminate at the open surfaces perpendicular to the $z$-axis and we can study the Majorana modes that can appear at the vortex ends in the topological phase. We compare these results, which are neither in the semi-classical, or infinitesimally weak-pairing limits, to the results studied in Ref. \onlinecite{hosur2011} which are in these limits.


\subsection{Periodic Vortex Rings in Vortex Lattices}
\label{sect:vortexring}

With periodic boundary conditions in all spatial directions, we cannot directly study the Majorana modes that might appear at the vortex ends. However, we can indirectly study them by identifying the vortex phase transition through a study of the low-energy modes along the vortex lines. As the chemical potential is tuned deeper into the band, the point where one of these modes becomes gapless signals the location of a critical point. For this geometry the magnetic unit cell sizes we use are  $l_x\times l_y
\times l_z=12\times 24\times 10$. We choose $N_x\times N_y=10 \times 5$ unit cells so that there are $100$ vortices in the vortex lattice.

\begin{figure}[t]
\epsfig{file=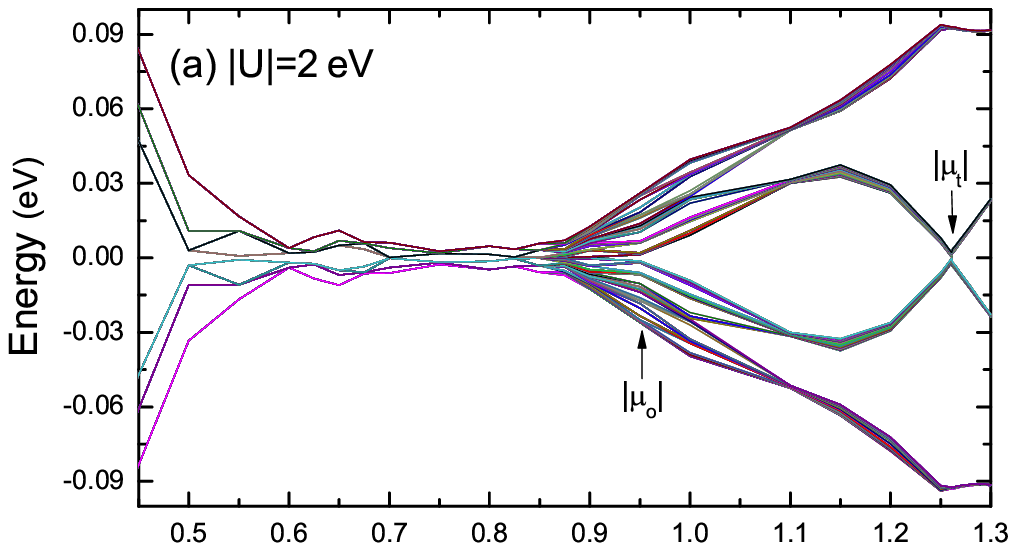,clip=0.7,width=1\linewidth,angle=0}
\epsfig{file=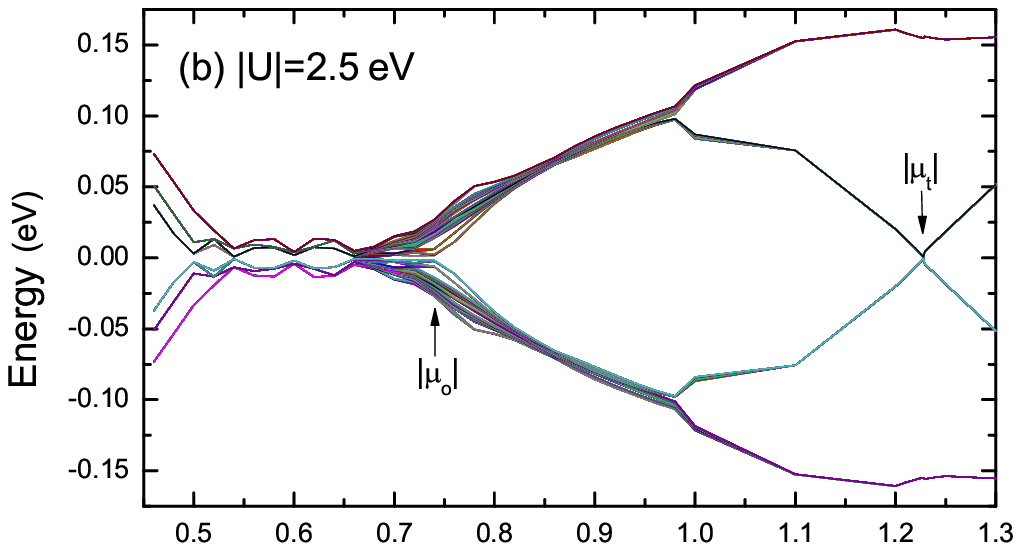,clip=0.7,width=1.01\linewidth,angle=0}
\epsfig{file=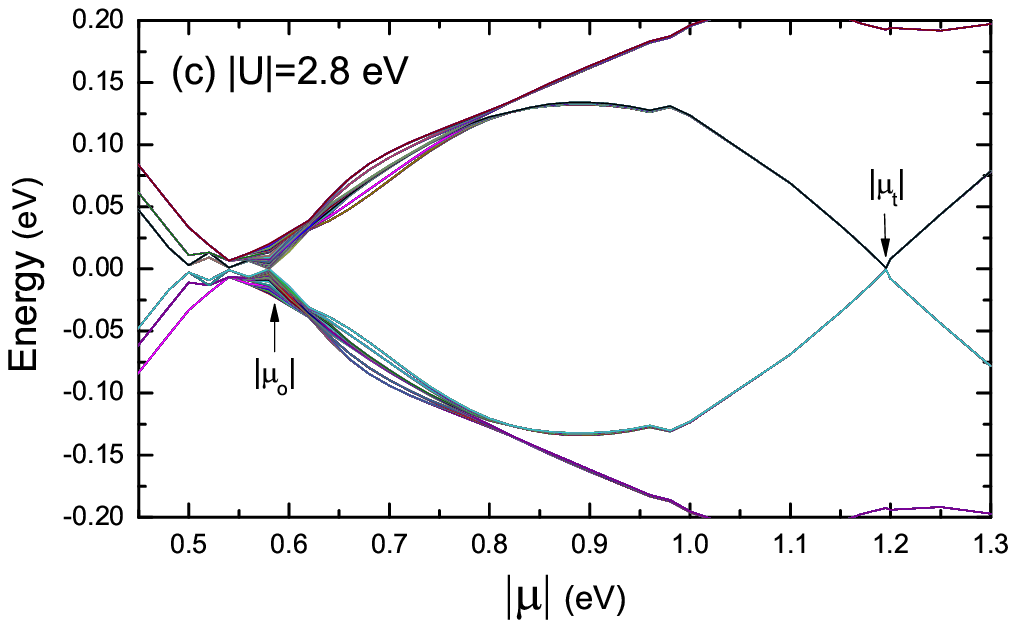,clip=0.7,width=1.01\linewidth,angle=0}
\caption{(color online). The energy spectra of the low-energy states vs $|\mu|$
for periodic boundary conditions along the $z$-axis (vortex rings) for (a) $|U|=2$ eV, (b) $|U|=2.5$ eV,
and (c) $|U|=2.8$ eV. The  systems have a $N_x\times N_y= 10\times 5$ (100 vortices) vortex lattice
and the size of the magnetic unit cell is $l_x\times l_y\times l_z=12a \times 24a\times
10a.$} \label{fig:vortexring}
\end{figure}


 In Fig. \ref{fig:vortexring}, we present the evolutions of the low energy states vs $|\mu|$ for different interaction strengths $|U|$. We can identify two distinctly different doping regimes. In the first regime, the chemical potential lies in the valence band but below a value we call $|\mu_o|$ which signals the onset of a well-formed superconducting gap discerned from our numerics. It should be noted that $|\mu_o|$ has no real intrinsic meaning (as it is strongly finite-size dependent) and only serves to indicate a common feature shared by all of our spectrum plots. Clearly, before the chemical potential hits the top of valence band there is no density of states to generate the superconducting gap, and all the states are gapped by the bulk insulating band gap. After the chemical potential hits the top of valence band, the superconducting pairing starts to form but it is exponentially small in magnitude. Comparing the pairing strengths without vortices  (\emph{i.e.} Fig. \ref{fig:periodiczTIdifferentU}) to the vortex lattice case in Fig. \ref{fig:vortexring}, our numerics show that in the vortex lattice the pairing is more poorly formed over a larger range of doping. That is, the states at the Fermi level remain gapless with no superconducting gap formation. 
 Note that $\mu_o$ decreases with increasing $|U|$ which indicates that this point is sensitive to the point where the exponentially suppressed superconducting gap would turn on.
 In this regime, any localized Majorana modes or low-energy vortex core states are difficult to distinguish from the extended gapless metallic states in the bulk. The details of this regime are dominated by strong finite-size effects. One hinderance is that for cases where only a tiny pairing potential would form it is numerically challenging to generate a convergent, self-consistent solution with vortices present. In the thermodynamic limit, we would expect to see a non-zero but exponentially small pairing gap as soon as the chemical potential hits the valence band. Here the situation is not so clear, and unfortunately, due to the computational limitations, we cannot glean a great deal of physical information from this regime except that it is not obvious that the picture of an ``adiabatic" continuation from the gapped insulating state immediately to a gapped superconducting state would be valid in a real sample. We will attempt to address this issue from another direction by studying a heterostructure geometry in Section \ref{sec:hetero} in which we can generate a convergent, vortex lattice solution by in homogeneously doping the system \emph{i.e.} high-doping on the surface and low-doping in the bulk. 

In the second distinct regime, once $|\mu|$ is tuned beyond
$|\mu_o|$, then significant s-wave pairing begins to develop.
Because of the particle-hole constraint of the BdG quasi-particle
spectrum, the energies appear in $\pm E$ pairs. The lowest energy
branches are nearly  $2\times N_x\times N_y$ fold degenerate. This
degeneracy clearly indicates that these states are in-gap vortex
states as there is essentially one for each vortex. As the chemical
potential is pushed more into the valence band, the lowest energy
branch approaches zero energy and at critical chemical potential
$|\mu_t|$, the particle and hole branches cross indicating the
location of the vortex phase transition. In the weak-pairing treatment
the critical chemical potential is independent of the value of the
attractive potential $|U|$ and if we repeat their analysis for our
choice of parameters, we find a weak-pairing estimate of
$|\mu_t|=1.35$ eV. In our case, as the interaction strength is quite
large we are not in the weak-pairing limit and the critical chemical
potential depends on the attractive potential. At $|U|=2$ eV, $2.5$
eV and $2.8$ eV, $|\mu_t| \simeq 1.26$ eV, $1.22$ eV and $1.2$ eV,
respectively. A stronger $|U|$ gives a smaller value of $|\mu_t|$
and it approaches to the weak paring limit as we decrease the
magnitude of interaction. Since the phenomenon survives the
weak-pairing limit it is possible that the vortex topological phase
transition could also be observed in a strong-pairing atomic limit which is
realizable in ultra-cold optical lattices\cite{beri2011}.

\begin{figure}[t]
\epsfig{file=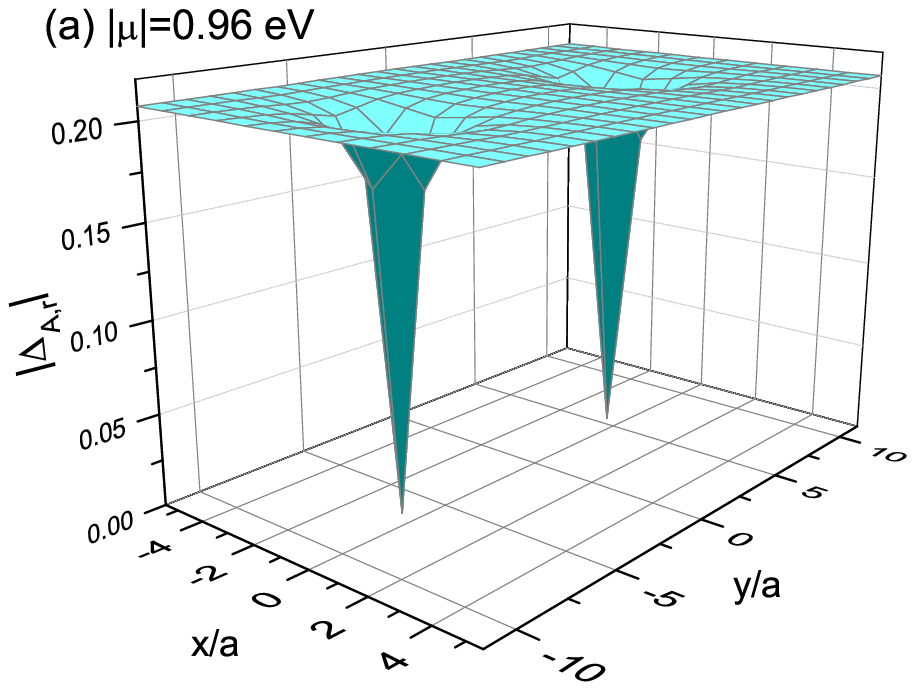,clip=0.7,width=0.7\linewidth,angle=0}
\epsfig{file=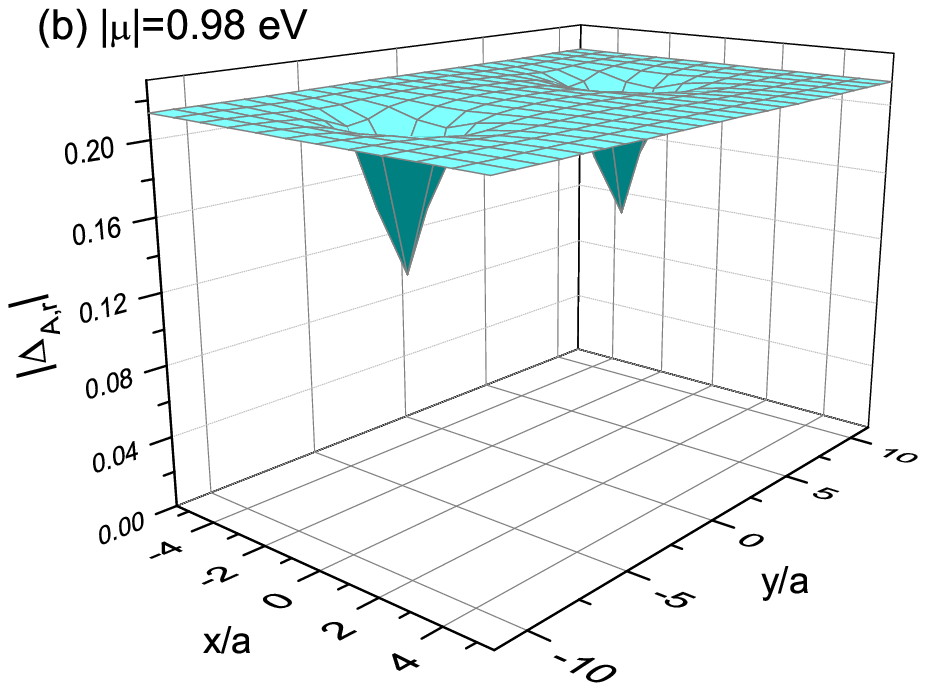,clip=0.7,width=0.7\linewidth,angle=0}
\epsfig{file=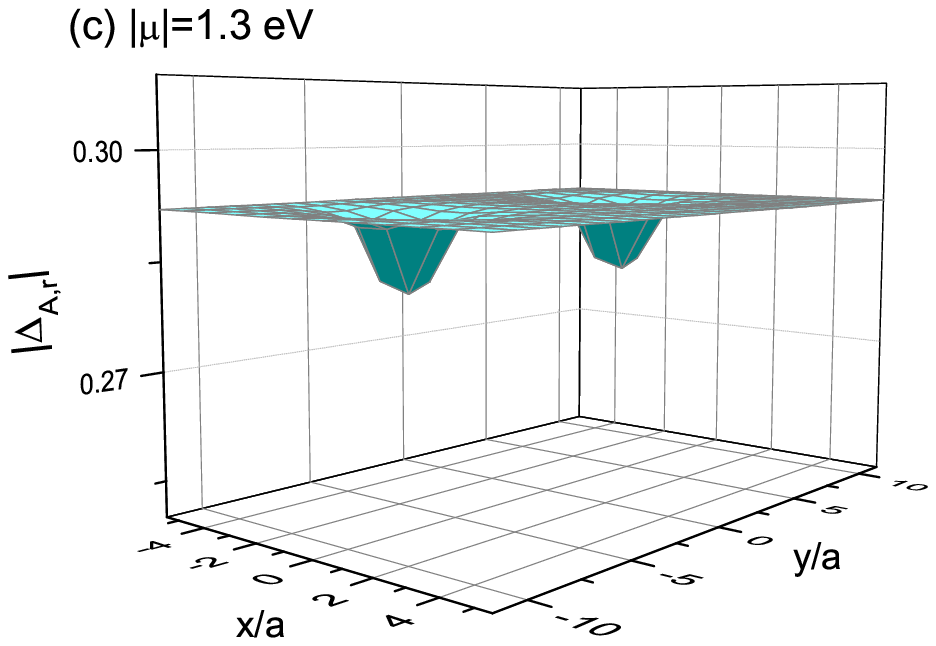,clip=0.7,width=0.7\linewidth,angle=0}
\caption{(color online). The spatial pairing order parameter profiles $|\Delta_{A,\vec{r}}|$ within a unit cell at $|U|=2.8$ eV and at (a) $|\mu|=0.96$ eV (b) $|\mu|=0.98$ eV and (c) $|\mu|=1.3$ eV.
The vertical axis represents the pairing order parameter magnitudes and the horizontal plane is $xy$ plane. Note the variation of the pairing order parameter magnitudes at
the unit cell center $\Delta_{\textrm{center}}=\Delta_{A,\vec{r} \in \textrm{unit cell center}}$ for different $\mu$: (a) $\Delta_{\textrm{center}}\simeq0$, (b) $\Delta_{\textrm{center}}\simeq0.12$, and (c) $\Delta_{\textrm{center}}\simeq0.28$. $\Delta_{\textrm{center}} \simeq 0$ indicates that the vortex core stays on-site, whereas $\Delta_{\textrm{center}}\ne 0$ means that the vortex core
moves off the lattice vertices and into the plaquette. We note that the profiles for $|\Delta_{B,\vec{r}}|$ look similar.}
\label{fig:vortexringpairingorder}
\end{figure}

In Fig. \ref{fig:vortexringpairingorder}, we show the
self-consistent vortex profiles in a single unit cell as a function
of $|\mu|$ for $|U|=2.8$ eV. It is evident that in all cases around
the vortex cores, the pairing order parameters are suppressed. Away from
the vortex cores the pairing order parameters are restored to 
$\Delta_{A}=0.21$, $0.213$ and $0.29$ which are roughly equal to the
corresponding bulk values at $|\mu|=0.96$ eV, $0.98$ eV and $1.3$ eV
respectively (c.f. Fig. \ref{fig:periodiczTIdifferentU}).
 In the
bulk superconducting TI, at larger $|\mu|$ stronger Cooper pairing
is induced, and the strong superconductivity leads to a shorter
coherence length $\xi_0$ ($\xi_0=\frac{\hbar v_F}{\pi
\Delta}$),\cite{degennes1966} and thus a smaller vortex size.
Therefore, in Fig. \ref{fig:vortexringpairingorder}(b) and (c), we
see flatter order parameter profiles. 
However, we find unusual behavior in these figures associated with
chemical potentials of $|\mu|=0.98$ eV and  $1.3$ eV. At these
chemical potentials, the vortices do not seem to be as well formed
as they are when $|\mu|=0.96$ eV. This is due to the numerical
discreteness in our simulations. In our system, because of the
non-trivial order parameter winding due to the vortex, there \emph{must} be a place where
the order parameter magnitude vanishes. One can see that in Fig.
\ref{fig:vortexringpairingorder} this only happens
for $|\mu|=0.96.$ What is happening is that the vortex core moves from being centered at a lattice vertex to the interior of a plaquette. The order parameter then vanishes in the plaquette interior
(which of course is not seen on our discrete lattice spatial sampling).
In fact, we find that at particular values of the
chemical potential it becomes energetically favorable for the vortex
to move its core off of a lattice vertex and into the center of a
plaquette.  This is seen in the energy spectra in Fig. \ref{fig:vortexring}
where we a kink in the spectrum a appears where
this vortex shift occurs, namely around $|\mu_k|=0.97$ eV. We believe this is simply an artifact of our numerical technique and does not represent any real physics.

\subsection{Open Vortex Lines in Vortex Lattices}

Next we turn to the case with open boundary conditions along the
$z$-direction such that the vortices terminate on the surfaces. This
setting is directly relevant for possible experiments where the
Majorana vortex modes are present at the end of vortex lines. Due to
the open boundaries, the self-consistent BdG calculations must be
performed in three spatial dimensions as we cannot exploit any
translation symmetry in $z$ direction. The magnetic unit cells are
$l_x\times l_y=8\times 16$ sized with $l_z$-layers, (usually
$l_z=6$) and there are $N_x\times N_y=40 \times 20$ magnetic unit
cells chosen so that we are simulating $1600$ vortices in the vortex
lattice. In this section, we only consider $|U|=2.8$ eV and
$|\mu|>|\mu_o|$ where the superconducting gap is formed and the
value of  $|\mu_o|$ is estimated from the periodic boundary
condition case. Here, we self-consistently determine the BdG
quasi-particle spectrum in the vortex lattice
state\cite{yasui1999,vafek2001,han2010}. For the square vortex
lattice with $N_x\times N_y$ magnetic unit cells, there are $N_x
N_y$ magnetic Bloch wavevectors $\vec{k}$ analogous to the
wavevectors in the Brillouin zone of a $\times N_x \times N_y$
square lattice. 

\begin{figure}[t]
\includegraphics[trim=6cm 0.5cm 0cm 0.5cm, clip=true, totalheight=0.52\textheight]{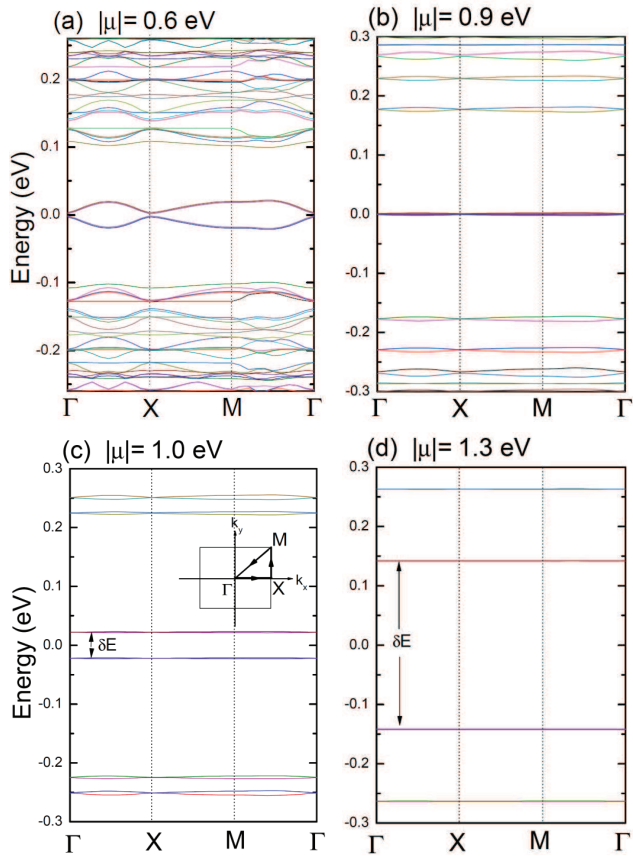}
\caption{(color online). The quasiparticle band structure for open
vortex lines in the bulk superconducting TI. The interaction
strength is chosen at $|U|=2.8$ eV and the chemical potentials are
(a) $|\mu|=0.6$ eV, (b) $|\mu|=0.9$ eV, (c) $|\mu|=1.0$ eV and (d)
$|\mu|=1.3$ eV. The inset in (c) denotes the magnetic Brillouin zone
for the square vortex lattice. The magnetic unit cell sizes are
$l_x\times l_y\times l_z=8a \times 16a\times 6a$.}
\label{fig:openvortexband}
\end{figure}

In Fig. \ref{fig:openvortexband}, we present the dispersion of
vortex modes at four different chemical potentials as has been done
previously for s-wave and d-wave
superconductors\cite{yasui1999,vafek2001}. The high symmetry points of the square lattice are at $\Gamma=(0,0)$,
$X=(\pi,0)$ and $M=(\pi,\pi)$, as indicated in the inset of Fig.
\ref{fig:openvortexband}(c). There are four low-energy
``Majorana" modes at each momentum which are contributed by the
two-vortices per cell and the two ends of each vortex line.  For a
single magnetic unit cell we would thus expect to see one Majorana
mode on the two ends of each of the two vortex lines giving rise to
a total of four vortices per cell. In this context, we put the word
Majorana in quotes because, strictly speaking, the low-lying energy
states only have true Majorana character if they are strictly at
zero-energy. In Fig. \ref{fig:openvortexband}(a), we study the
quasiparticle bands at  $|\mu|=0.6$ eV and we find that the vortex
modes are clearly dispersing. Although the superconducting gap is
formed, it remains small and the vortex modes of different vortices
on the \emph{same} surface can tunnel laterally and hybridize which leads to the dispersion of
the vortex core states.  As we increase the doping level, the
lowest energy quasiparticle band flattens as is clear in the
dispersion plot for $|\mu|=0.9$ eV in Fig.
\ref{fig:openvortexband}(b). This happens because of the increasing
bulk superconducting gap, indicated in Fig. \ref{fig:vortexring}(c).
The vortex core size shrinks which leads to smaller overlap of the
modes localized in different vortices and suppresses the
quasiparticle dispersion. This effect (\emph{i.e.} increase of
superconducting gap by increasing the doping) stabilizes the
Majorana modes. The low-energy, flat quasi-particle bands contain $4N_xN_y$ nearly-degenerate states coming from the $2N_x N_y$ vortex Majorana modes on the two distinct surfaces.

In Fig. \ref{fig:openvortexband}, we see two clear gap-like behaviors. One type in Fig. \ref{fig:openvortexband}(a) shows gaps at low-energy but with strong dispersion while Fig. \ref{fig:openvortexband}(c) and (d) show clear gaps but with flat dispersion. For the flat-dispersing cases we studied the dependence of the energy splitting, $\delta E$, on the sample thickness. An exponential dependence would indicate that the dispersionless gap is a result of the hybridization of the modes at the end of the vortices between two surfaces.
\begin{figure}[t]
\epsfig{file=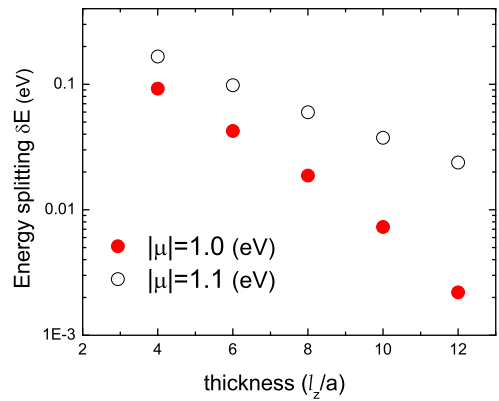,clip=0.7,width=0.75\linewidth,angle=0}
\caption{(color online). The energy splitting $\delta E$ vs thickness $l_z$. The magnetic unit cells are $8 \times 16 \times l_z$ at $|\mu|=1$ eV and $|\mu|=1.1$ eV.
The Hubbard interaction is $|U|=2.8$ eV.} \label{fig:bulkEvsthickness}
\end{figure}
\noindent Fig. \ref{fig:bulkEvsthickness} shows the energy splitting $\delta E$ has an exponential decreasing relation with the thickness $l_z$ described as\cite{kitaev2001}
\bea \delta E \propto  e^{-\frac{l_z}{\xi_m}},\eea
where $\xi_m$ denotes the characteristic decay length for the Majorana modes.
A smaller $\xi_m$ means more localized Majorana bound states.
By linear fitting from Fig. \ref{fig:bulkEvsthickness}, the characteristic length at $|\mu|=1$ eV is $\xi_m \simeq 5.46a$ and at $|\mu|=1.1$ eV is $\xi_m \simeq 9.49a$. This suggests that the Majorana modes are still exponentially localized on the surface even though there is a gap in our finite-size numerics.
Therefore, although the Majorana modes may tunnel to the opposite surface,
 in the thermodynamic limit ($l_z \to \infty$), the Majorana modes are still be bound to the surface. Although we do not show it here, we note that this is not the case for $\vert \mu\vert =1.3$ eV where the gap does not decrease exponentially, which is expected since this is the trivial regime where it should have a power-law decay with inverse thickness due to finite-size splitting.

\begin{figure*}[t]
\centering
\includegraphics[trim=2cm 6cm 0cm 4cm, clip=true, totalheight=0.315\textheight]{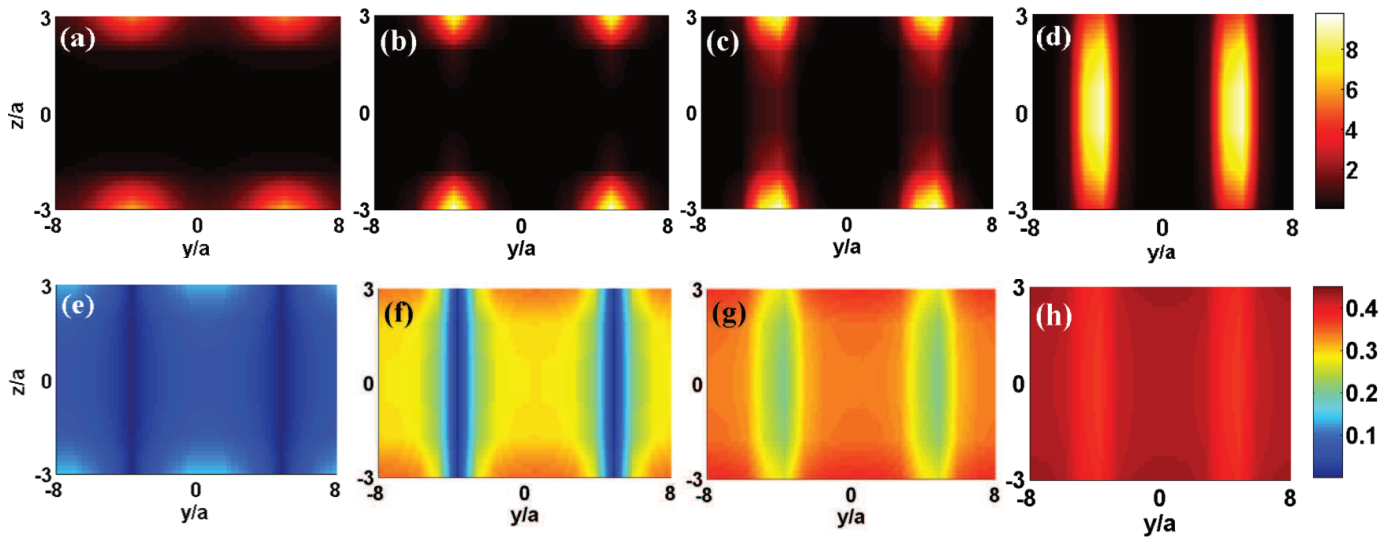}
\caption{(color online). Spatial slices (side-views in the $yz$ plane
at $x=\pm2a$) of the probability density for the Majorana modes and order parameter density in the
bulk superconducting TI at different $\mu$. Upper panels: (a)-(d)
shows the evolution of the Majorana mode distributions. Brighter
regions represent the higher probability density. Lower
panels: (e)-(h) shows the distribution of pairing order parameters
($|\Delta_A+\Delta_B|$). The chemical potentials are $|\mu|=0.6$ eV
in (a) and (e), $|\mu|=0.9$ eV in (b) and (f), $|\mu|=1$ eV in (c)
and (g), and $|\mu|=1.3$ eV in (d) and (h), respectively.}
\label{fig:openvortexsideview}
\end{figure*}

The nature of the Majorana modes may be further illustrated by
studying real space probability distributions of the in-gap modes.
Fig. \ref{fig:openvortexsideview}(a)-(d) depict side-view spatial slices of the probability density for the
lowest energy modes and Figs. \ref{fig:openvortexsideview}(e)-(h)
show the  order parameter distributions in real space. The
plots are cut on the $yz$ surface at $x=\pm 2a$, where the vortex
cores are approximately located. The Majorana modes (indicated by
bright regions) are observed and localized around the vortex cores
close to the surfaces in Figs. \ref{fig:openvortexsideview}(a) and
(b). However, the Majorana mode in Fig.
\ref{fig:openvortexsideview}(a) spreads more widely along the
surface than that in Fig. \ref{fig:openvortexsideview}(b). This
shows that at $|\mu|=0.6$ eV neighboring vortices have larger
overlap than that at $|\mu|=0.9$ eV, which corroborates with our
quasiparticle spectra that indicate stronger dispersion for the
former case as shown in Fig. \ref{fig:openvortexband}(a) due to the
intra-surface hybridization resulting from the increased lateral
overlap of the Majorana modes. It is also interesting to see that
around $|\mu|=0.9$ eV, the mini-gap size of the vortex lines is
maximum, [see Fig. \ref{fig:vortexring}(c)] which is where and Fig.
\ref{fig:openvortexsideview}(f) shows strong, straight-line vortex
structures.

At first, further increases in the chemical potential flattens the dispersion and strengthens the localization of the Majorana modes. However, further increases in the chemical potential lead to another tunneling mechanism for the Majorana modes. As we have already shown for periodic vortex rings (e.g see Fig. \ref{fig:vortexring}(c) as $|\mu|>1$) the mini-gap of the vortex core states along the vortex line eventually begins to decrease as the critical point is approached. For open-boundary conditions this leads to increased inter-surface hybridization of the modes at the
two ends of the vortex lines. This results in the formation of gaps due to  Majorana mode annihilation on opposite surfaces (for thin samples) and in Fig. \ref{fig:openvortexband}(c) we can see that there exists a $\delta E$ splitting in
the Majorana modes. As mentioned and shown in Fig. \ref{fig:bulkEvsthickness} $\delta E$ decreases exponentially in the thickness of the sample.  
An important feature to note is that, as is clear from the dispersion for
$|\mu|=0.9$ eV,  even though the gap increases the
bandwidth of quasiparticles decreases and gets flatter. This is an indication that intra-surface tunneling is weakening (no 2D hopping on the same surface) and that inter-surface tunneling is becoming stronger.
We can see this in Fig. \ref{fig:openvortexsideview}(c), where the Majorana bound states begin to leak to the opposite surface. Furthermore, in Fig. \ref{fig:openvortexsideview} (d), in which case the system is topologically trivial, the lowest energy modes, which are no longer Majorana in nature and gapped by the vortex mini-gap of order $\Delta^2/\mu$, completely penetrate through the bulk at $|\mu|=1.3$ eV and lie along the vortex lines.

\subsection{Superconductor-Topological Insulator Heterostructure}\label{sec:hetero}
As mentioned in the introduction,  another method to realize the
Majorana modes is through the proximity effect of a topological
insulator and an s-wave superconductor. Using our self-consistent BdG
method, we can also study this geometry. By modeling such a structure by an inhomogenous doping level we can directly
address the effect of the penetration of superconducting gap into the
bulk and the self-consistent formation of Majorana bound states in vortices. 
 We imagine a similar proximity-induced
superconductivity as was first
suggested by Fu and Kane\cite{fu2008}. We choose an inhomogenous system
where $\mu(\vec{r})$ in Eq. (\ref{eq:newbdg}) is
layer-dependent. We choose the surface chemical potential
$\mu(\vec{r})=\mu_S=-1$ eV, and the chemical potential
$\mu(\vec{r})=\mu_B=-0.55$ eV in other bulk layers so that both the
bulk and surface have non-zero density of states.
We investigate the
case where superconductivity is induced primarily on the surfaces by
turning on the same attractive interaction across the entire sample.
The inhomogeneous $\mu(\vec{r})$ will generate a much stronger order
parameter on the surfaces than in the bulk due to the large
difference in chemical potentials. Again we choose a uniform magnetic field along the $z$-direction
which generates the vortex lattice.
\begin{figure}[t]
\epsfig{file=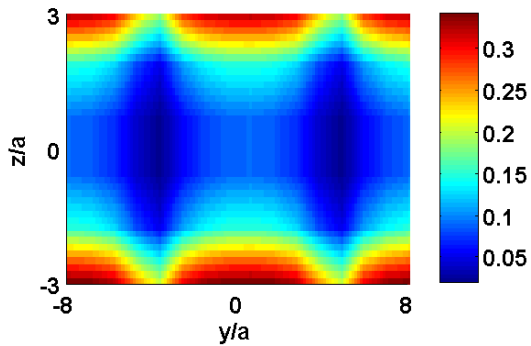,clip=0.7,width=0.8\linewidth,angle=0}
\caption{(color online). The spatial side view of the pairing order parameter distribution of a six-layer the s-wave/TI heterostructure.
The interaction strength is chosen at $|U|=2.8$ eV, and the surface and bulk chemical potentials are $|\mu_S|=1$ eV and $|\mu_B|=0.55$ eV, respectively.
The dark blue tubes indicate the region that pairing is suppressed and form vortex lines. The magnetic unit cell sizes are $8a \times 16a\times
6a$} \label{fig:heteropairing}
\end{figure}

\begin{figure}[t]
\includegraphics[trim=5cm 0cm 0cm 1cm, clip=true, totalheight=0.52\textheight]{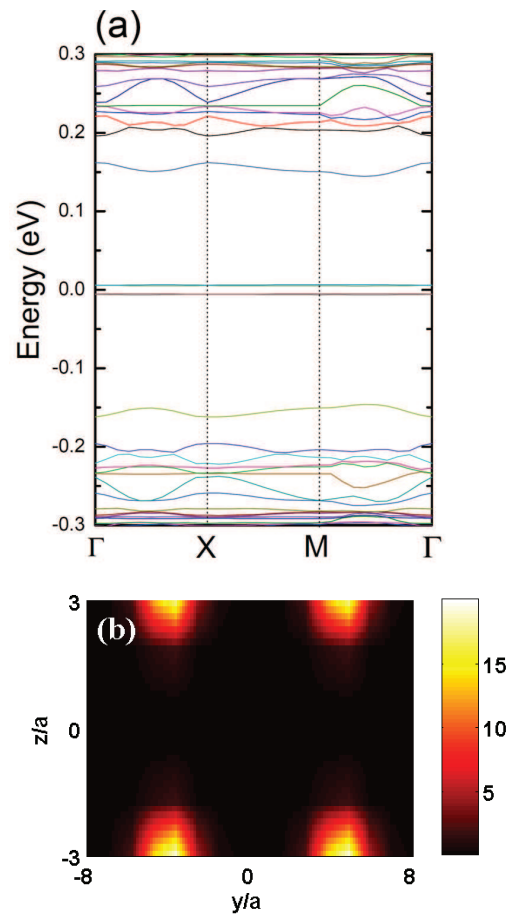}
\caption{(color online). (a) The quasiparticle band spectrum for six-layer the $s$-wave/TI heterostructure. (b) The spatial side view of the Majorana modes. The interaction strength is chosen at $|U|=2.8$ eV and $|\mu_S|=1$ eV and $|\mu_B|=0.55$ eV.} \label{fig:heterbandzeromode}
\end{figure}

There is another reason to consider this system beyond simply the presence of Majorana fermions. 
One of the major obstacles in our
bulk calculation is the non-convergence of a stable vortex solution
when the order parameter magnitude is very small. 
As mentioned,
when $|\mu_v|<|\mu| < |\mu_o|$ despite the presence of gapless
electrons at the Fermi-level, the density of states is not large enough to form a
sizable superconducting gap (at least for system sizes we consider) and the doped
TI remains a gapless metal. 
We can counter-act
this problem by using the superconductor-TI heterostructure geometry which acts to pin the vortices with strong superconductivity at the surface (highly-doped region)
thereby stabilizing the solution.  
  Fig. \ref{fig:heteropairing} shows the spatial side view of
the resulting self-consistent order parameter profile of a six-layer heterostructure. We see no evidence of
superconductivity in the bulk and roughly uniform
superconductivity in the surface which is interrupted
near the vortices. 
The resulting calculation in Fig.
\ref{fig:heterbandzeromode}b shows that Majorana surface modes still
remain even though the superconducting order parameter in the bulk
is exponentially small compared to the surface. The six-layer heterostructure can roughly
approximate the case of two vortices existing in a four-layer bulk TI 
which is \emph{uniformly} doped with $|\mu|=0.55$ eV. This was a
region of interest that we could not access in our bulk calculation
due to finite-size complications and which we can, admittedly only roughly, learn about by stabilizing the vortex solution using higher surface doping.

In Fig. \ref{fig:heterbandzeromode}(a), we present the
quasi-particle band spectrum for the superconductor-TI
heterostructure with a square vortex lattice. Within the superconducting gap, there exist two
prominent low-energy modes which are doubly degenerate whose
energies are split away from zero energy. Although not shown here,
we find that the energy splitting $\delta E$ also has an exponential
decay with increasing sample thickness $l_z.$ This indicates that the
low-energy modes are exponentially localized at the superconducting
surface and the Majorana fermions can stably reside at the surfaces
in the thermodynamic limit, i.e. $l_z \to \infty$. This is indicative that 
that the vortices in the low-doping regime
($|\mu_v|<|\mu|<|\mu_o|$) also support Majorana
fermions in the bulk superconducting TI. In Fig.
\ref{fig:heterbandzeromode}(b) we show the probability density of
the low energy modes and see the tight localization of the resulting
modes on the surface as one would expect for Majorana modes formed
within a well-formed superconducting gap.

\section{Vortex Majorana Modes at Finite Temperatures}\label{sec:finiteT}
\begin{figure}[t]
\includegraphics[width=7cm]{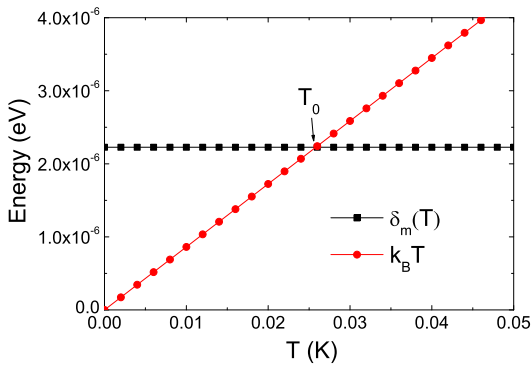}
\caption{(color online) The comparison between the midgap sizes
$\delta_m(T)$ and $k_B T$. The intersection occurs at $T=T_0\sim
0.025$ K indicating that as $T<T_0$, the Majorana modes can stably
exist on the surface of the doped topological insulators.}
\label{fig:midgapvsT}
\end{figure}

In our previous analysis, we have neglected the role of temperature in our analysis. In this section, we provide a rough estimate of the temperature at which one could observe the Majorana fermions experimentally.
In the BCS theory the temperature dependence of the gap size is determined through:\cite{degennes1966}
\begin{equation}
\label{eq:finiteTgap}
 \frac{1}{N(0)V}=\int_0^{\hbar \omega_c} \frac{\tanh{\left[ \frac{\sqrt{\xi^2+\Delta(T)^2}}{2k_BT}   \right]}}{\sqrt{\xi^2+\Delta(T)^2}} d\xi.
\end{equation}
Here $N(0)$ is the density of states at the Fermi-level, $U$ is the interaction coupling, and $\omega_c$ is the Debye frequency. The finite temperature gap  $\Delta(T)$ may only be determined numerically. The combination of $N(0)U$ can be estimated from the critical temperature, $T_c$, and the Debye frequency, $\omega_c$, for Cu-doped Bi$_2$Se$_3$  via:

\begin{equation}
\label{eq:goV}
N(0)U=\frac{-1}{\ln{\left[\frac{k_B T_c}{1.13 \hbar \omega_c}\right]}}.
\end{equation}

For Bi$_2$Se$_3$, the critical temperature is $T_c=3.8$K,\cite{hor2010,hor2011} and the Debye temperature $\hbar \omega_c/k_B$ is 180 K\cite{shoemake1969}. With $N(0)U$ determined from  Eq. (\ref{eq:goV}), one can calculate the temperature dependence of the gap numerically using Eq. (\ref{eq:finiteTgap}). To observe the Majorana modes at finite temperature, the mini-gap size of the vortex lines $\delta_m(T)$ should be stable against the thermal fluctuations: $\delta_m(T) < k_B T$. The temperature at which this occurs may be estimated as
\begin{equation}
 T_0= \frac{\delta_m(T)}{k_B}=\frac{\pi \Delta(T)^2}{2k_B \delta \varepsilon_F},
\end{equation}
\noindent where $\delta \varepsilon_F=\vert \mu-\mu_b\vert$ where  $\mu_b$
denotes the bulk band edge. When $T< T_0$, the Majorana modes can
stably exist on the surfaces and can be detected experimentally. The
numerical result is shown in Fig. \ref{fig:midgapvsT}. From Ref.
\onlinecite{wray2010}, in Bi$_2$Se$_3$, $\delta
\varepsilon_F\sim0.25$ eV which results in an estimate for the
critical temperature for the observation of Majorana modes to be
$T_0\sim 0.025$ K. Therefore, we can provide a rough estimate that
at $T \lesssim 0.025$ K, the Majorana modes can stably exist on the
surface of the doped topological insulators and may be detectable
experimentally. This number is quite small and indicates one would
need to optimize materials properties in order to hope for
observation. The results for Heusler materials or materials with
similar electronic structure to bulk HgTe may provide more promising
alternatives\cite{chiughaemihughes} due to the differences in the
sustainable levels of doping.

\section{conclusion}
\label{sect:conclusion}

In summary, we performed self-consistent Bogoliubov-de-Gennes calculations to study properties of vortices in doped topological insulators that become superconducting. Through the use of our numerics, we studied the physics of Majorana fermions in vortex lattices beyond the strict weak-coupling limit, and the resulting vortex phase transitions between a topological and trivial state.  We have shown that the quasi-particle band spectra offers evidence that there exists an optimal regime in chemical potential where the Majorana fermions can stably reside even in a finite thickness system. There also exists other regimes where the Majorana fermions do not stably exist on the system surfaces because of intra- and inter-surface hybridization between the vortex modes. Furthermore, we also showed that, through the use of the analogous s-wave-TI heterostructure,  that TIs with bulk superconductivity containing finite carrier density but insufficient superconducting pairing strength can host Majorana fermions on the surface. Similar to the bulk superconducting case, the Majorana modes can also leak into the bulk and annihilate with the other surface. However, the tunneling behavior exhibits the usual exponential decay with thickness and we conclude that the Majorana fermions can survive for thick samples. Unfortunately, the simple estimates we made for a viable temperature range in which Majorana modes may be observed indicate that superconducting Cu-Bi$_2$Se$_3$ may not provide a good candidate even if the doping level can be tuned to the topological vortex phase.

\acknowledgements HHH is grateful to helpful discussions with C.-K
Chiu. PG is thankful for useful discussions with E. Fradkin and P.
Goldbart and support under the grant NSF DMR-1064319.
This work was partially supported in part by the National Science
Foundation under Grant NSF-OCI 1053575. TLH  acknowledges support from U.S. DOE, Office of Basic Energy Sciences, Division of Materials Sciences and Engineering under Award DE-FG02-07ER46453. 
MJG and HHH acknowledge support
from the AFOSR under grant FA9550-10-1-0459. We acknowledge support
from the Center for Scientific Computing at the CNSI and MRL: an NSF
MRSEC (DMR-1121053) and NSF CNS-0960316.


\bibliography{TI}

\begin{thebibliography}{53}%
\makeatletter
\providecommand \@ifxundefined [1]{%
 \@ifx{#1\undefined}
}%
\providecommand \@ifnum [1]{%
 \ifnum #1\expandafter \@firstoftwo
 \else \expandafter \@secondoftwo
 \fi
}%
\providecommand \@ifx [1]{%
 \ifx #1\expandafter \@firstoftwo
 \else \expandafter \@secondoftwo
 \fi
}%
\providecommand \natexlab [1]{#1}%
\providecommand \enquote  [1]{``#1''}%
\providecommand \bibnamefont  [1]{#1}%
\providecommand \bibfnamefont [1]{#1}%
\providecommand \citenamefont [1]{#1}%
\providecommand \href@noop [0]{\@secondoftwo}%
\providecommand \href [0]{\begingroup \@sanitize@url \@href}%
\providecommand \@href[1]{\@@startlink{#1}\@@href}%
\providecommand \@@href[1]{\endgroup#1\@@endlink}%
\providecommand \@sanitize@url [0]{\catcode `\\12\catcode `\$12\catcode
  `\&12\catcode `\#12\catcode `\^12\catcode `\_12\catcode `\%12\relax}%
\providecommand \@@startlink[1]{}%
\providecommand \@@endlink[0]{}%
\providecommand \url  [0]{\begingroup\@sanitize@url \@url }%
\providecommand \@url [1]{\endgroup\@href {#1}{\urlprefix }}%
\providecommand \urlprefix  [0]{URL }%
\providecommand \Eprint [0]{\href }%
\providecommand \doibase [0]{http://dx.doi.org/}%
\providecommand \selectlanguage [0]{\@gobble}%
\providecommand \bibinfo  [0]{\@secondoftwo}%
\providecommand \bibfield  [0]{\@secondoftwo}%
\providecommand \translation [1]{[#1]}%
\providecommand \BibitemOpen [0]{}%
\providecommand \bibitemStop [0]{}%
\providecommand \bibitemNoStop [0]{.\EOS\space}%
\providecommand \EOS [0]{\spacefactor3000\relax}%
\providecommand \BibitemShut  [1]{\csname bibitem#1\endcsname}%
\let\auto@bib@innerbib\@empty
\bibitem [{\citenamefont {Hosur}\ \emph {et~al.}(2011)\citenamefont {Hosur},
  \citenamefont {Ghaemi}, \citenamefont {Mong},\ and\ \citenamefont
  {Vishwanath}}]{hosur2011}%
  \BibitemOpen
  \bibfield  {author} {\bibinfo {author} {\bibfnamefont {P.}~\bibnamefont
  {Hosur}}, \bibinfo {author} {\bibfnamefont {P.}~\bibnamefont {Ghaemi}},
  \bibinfo {author} {\bibfnamefont {R.~S.~K.}\ \bibnamefont {Mong}}, \ and\
  \bibinfo {author} {\bibfnamefont {A.}~\bibnamefont {Vishwanath}},\ }\href
  {\doibase 10.1103/PhysRevLett.107.097001} {\bibfield  {journal} {\bibinfo
  {journal} {Phys. Rev. Lett.}\ }\textbf {\bibinfo {volume} {107}},\ \bibinfo
  {pages} {097001} (\bibinfo {year} {2011})}\BibitemShut {NoStop}%
\bibitem [{\citenamefont {Majorana}(1937)}]{majorana1937}%
  \BibitemOpen
  \bibfield  {author} {\bibinfo {author} {\bibfnamefont {E.}~\bibnamefont
  {Majorana}},\ }\href@noop {} {\bibfield  {journal} {\bibinfo  {journal}
  {Niovo Cimento}\ }\textbf {\bibinfo {volume} {5}},\ \bibinfo {pages} {171}
  (\bibinfo {year} {1937})}\BibitemShut {NoStop}%
\bibitem [{\citenamefont {Ivanov}(2001)}]{ivanov2001}%
  \BibitemOpen
  \bibfield  {author} {\bibinfo {author} {\bibfnamefont {D.~A.}\ \bibnamefont
  {Ivanov}},\ }\href@noop {} {\bibfield  {journal} {\bibinfo  {journal} {Phys.
  Rev. Lett.}\ }\textbf {\bibinfo {volume} {86}},\ \bibinfo {pages} {268}
  (\bibinfo {year} {2001})}\BibitemShut {NoStop}%
\bibitem [{\citenamefont {Kitaev}(2001)}]{kitaev2001}%
  \BibitemOpen
  \bibfield  {author} {\bibinfo {author} {\bibfnamefont {A.~Y.}\ \bibnamefont
  {Kitaev}},\ }\href@noop {} {\bibfield  {journal} {\bibinfo  {journal} {Phys.
  Usp.}\ }\textbf {\bibinfo {volume} {44}},\ \bibinfo {pages} {131} (\bibinfo
  {year} {2001})}\BibitemShut {NoStop}%
\bibitem [{\citenamefont {Wilczek}(2009)}]{wilczek2009}%
  \BibitemOpen
  \bibfield  {author} {\bibinfo {author} {\bibfnamefont {F.}~\bibnamefont
  {Wilczek}},\ }\href@noop {} {\bibfield  {journal} {\bibinfo  {journal} {Nat.
  Phys.}\ }\textbf {\bibinfo {volume} {5}},\ \bibinfo {pages} {614} (\bibinfo
  {year} {2009})}\BibitemShut {NoStop}%
\bibitem [{\citenamefont {Hughes}(2011)}]{hughes2011}%
  \BibitemOpen
  \bibfield  {author} {\bibinfo {author} {\bibfnamefont {T.~L.}\ \bibnamefont
  {Hughes}},\ }\href@noop {} {\bibfield  {journal} {\bibinfo  {journal}
  {Physics}\ }\textbf {\bibinfo {volume} {4}},\ \bibinfo {pages} {67} (\bibinfo
  {year} {2011})}\BibitemShut {NoStop}%
\bibitem [{\citenamefont {Das~Sarma}\ \emph {et~al.}(2006)\citenamefont
  {Das~Sarma}, \citenamefont {Nayak},\ and\ \citenamefont
  {Tewari}}]{dassarma2006}%
  \BibitemOpen
  \bibfield  {author} {\bibinfo {author} {\bibfnamefont {S.}~\bibnamefont
  {Das~Sarma}}, \bibinfo {author} {\bibfnamefont {C.}~\bibnamefont {Nayak}}, \
  and\ \bibinfo {author} {\bibfnamefont {S.}~\bibnamefont {Tewari}},\
  }\href@noop {} {\bibfield  {journal} {\bibinfo  {journal} {Phys. Rev. B}\
  }\textbf {\bibinfo {volume} {73}},\ \bibinfo {pages} {220502} (\bibinfo
  {year} {2006})}\BibitemShut {NoStop}%
\bibitem [{\citenamefont {Read}\ and\ \citenamefont {Green}(2000)}]{read2000}%
  \BibitemOpen
  \bibfield  {author} {\bibinfo {author} {\bibfnamefont {N.}~\bibnamefont
  {Read}}\ and\ \bibinfo {author} {\bibfnamefont {D.}~\bibnamefont {Green}},\
  }\href@noop {} {\bibfield  {journal} {\bibinfo  {journal} {Phys. Rev. B}\
  }\textbf {\bibinfo {volume} {61}},\ \bibinfo {pages} {10267} (\bibinfo {year}
  {2000})}\BibitemShut {NoStop}%
\bibitem [{\citenamefont {Nayak}\ \emph {et~al.}(2008)\citenamefont {Nayak},
  \citenamefont {Simon}, \citenamefont {Stern}, \citenamefont {Freedman},\ and\
  \citenamefont {Das~Sarma}}]{nayak2008}%
  \BibitemOpen
  \bibfield  {author} {\bibinfo {author} {\bibfnamefont {C.}~\bibnamefont
  {Nayak}}, \bibinfo {author} {\bibfnamefont {S.~H.}\ \bibnamefont {Simon}},
  \bibinfo {author} {\bibfnamefont {A.}~\bibnamefont {Stern}}, \bibinfo
  {author} {\bibfnamefont {M.}~\bibnamefont {Freedman}}, \ and\ \bibinfo
  {author} {\bibfnamefont {S.}~\bibnamefont {Das~Sarma}},\ }\href@noop {}
  {\bibfield  {journal} {\bibinfo  {journal} {Rev. Mod. Phys.}\ }\textbf
  {\bibinfo {volume} {80}},\ \bibinfo {pages} {1083} (\bibinfo {year}
  {2008})}\BibitemShut {NoStop}%
\bibitem [{\citenamefont {Willett}\ \emph {et~al.}(1987)\citenamefont
  {Willett}, \citenamefont {Eisenstein}, \citenamefont {Stormer}, \citenamefont
  {Tsui}, \citenamefont {Gossard},\ and\ \citenamefont
  {English}}]{willett1987}%
  \BibitemOpen
  \bibfield  {author} {\bibinfo {author} {\bibfnamefont {R.}~\bibnamefont
  {Willett}}, \bibinfo {author} {\bibfnamefont {J.~P.}\ \bibnamefont
  {Eisenstein}}, \bibinfo {author} {\bibfnamefont {H.~L.}\ \bibnamefont
  {Stormer}}, \bibinfo {author} {\bibfnamefont {D.~C.}\ \bibnamefont {Tsui}},
  \bibinfo {author} {\bibfnamefont {A.~C.}\ \bibnamefont {Gossard}}, \ and\
  \bibinfo {author} {\bibfnamefont {J.~H.}\ \bibnamefont {English}},\
  }\href@noop {} {\bibfield  {journal} {\bibinfo  {journal} {Phys. Rev. Lett.}\
  }\textbf {\bibinfo {volume} {59}},\ \bibinfo {pages} {1776} (\bibinfo {year}
  {1987})}\BibitemShut {NoStop}%
\bibitem [{\citenamefont {Moore}\ and\ \citenamefont {Read}(1991)}]{moore1991}%
  \BibitemOpen
  \bibfield  {author} {\bibinfo {author} {\bibfnamefont {G.}~\bibnamefont
  {Moore}}\ and\ \bibinfo {author} {\bibfnamefont {N.}~\bibnamefont {Read}},\
  }\href@noop {} {\bibfield  {journal} {\bibinfo  {journal} {Nucl. Phys. B}\
  }\textbf {\bibinfo {volume} {360}},\ \bibinfo {pages} {362} (\bibinfo {year}
  {1991})}\BibitemShut {NoStop}%
\bibitem [{\citenamefont {Radu}\ \emph {et~al.}(2008)\citenamefont {Radu},
  \citenamefont {Miller}, \citenamefont {Marcus}, \citenamefont {Kastner},
  \citenamefont {Pfeiffer},\ and\ \citenamefont {West}}]{radu2008}%
  \BibitemOpen
  \bibfield  {author} {\bibinfo {author} {\bibfnamefont {I.~P.}\ \bibnamefont
  {Radu}}, \bibinfo {author} {\bibfnamefont {J.~B.}\ \bibnamefont {Miller}},
  \bibinfo {author} {\bibfnamefont {C.~M.}\ \bibnamefont {Marcus}}, \bibinfo
  {author} {\bibfnamefont {M.~A.}\ \bibnamefont {Kastner}}, \bibinfo {author}
  {\bibfnamefont {L.~N.}\ \bibnamefont {Pfeiffer}}, \ and\ \bibinfo {author}
  {\bibfnamefont {K.~W.}\ \bibnamefont {West}},\ }\href@noop {} {\bibfield
  {journal} {\bibinfo  {journal} {Science}\ }\textbf {\bibinfo {volume}
  {320}},\ \bibinfo {pages} {899} (\bibinfo {year} {2008})}\BibitemShut
  {NoStop}%
\bibitem [{\citenamefont {Kitaev}(2003)}]{kitaev2003}%
  \BibitemOpen
  \bibfield  {author} {\bibinfo {author} {\bibfnamefont {A.~Y.}\ \bibnamefont
  {Kitaev}},\ }\href@noop {} {\bibfield  {journal} {\bibinfo  {journal} {Ann.
  Phys.}\ }\textbf {\bibinfo {volume} {303}},\ \bibinfo {pages} {2} (\bibinfo
  {year} {2003})}\BibitemShut {NoStop}%
\bibitem [{\citenamefont {Cheng}\ \emph {et~al.}(2009)\citenamefont {Cheng},
  \citenamefont {Lutchyn}, \citenamefont {Galitski},\ and\ \citenamefont
  {Das~Sarma}}]{cheng2009}%
  \BibitemOpen
  \bibfield  {author} {\bibinfo {author} {\bibfnamefont {M.}~\bibnamefont
  {Cheng}}, \bibinfo {author} {\bibfnamefont {R.~M.}\ \bibnamefont {Lutchyn}},
  \bibinfo {author} {\bibfnamefont {V.}~\bibnamefont {Galitski}}, \ and\
  \bibinfo {author} {\bibfnamefont {S.}~\bibnamefont {Das~Sarma}},\ }\href
  {\doibase 10.1103/PhysRevLett.103.107001} {\bibfield  {journal} {\bibinfo
  {journal} {Phys. Rev. Lett.}\ }\textbf {\bibinfo {volume} {103}},\ \bibinfo
  {pages} {107001} (\bibinfo {year} {2009})}\BibitemShut {NoStop}%
\bibitem [{\citenamefont {Jang}\ \emph {et~al.}(2011)\citenamefont {Jang},
  \citenamefont {Ferguson}, \citenamefont {Vakaryuk}, \citenamefont {Budakian},
  \citenamefont {Chung}, \citenamefont {Goldbart},\ and\ \citenamefont
  {Maeno}}]{jang2011}%
  \BibitemOpen
  \bibfield  {author} {\bibinfo {author} {\bibfnamefont {J.}~\bibnamefont
  {Jang}}, \bibinfo {author} {\bibfnamefont {D.~G.}\ \bibnamefont {Ferguson}},
  \bibinfo {author} {\bibfnamefont {V.}~\bibnamefont {Vakaryuk}}, \bibinfo
  {author} {\bibfnamefont {R.}~\bibnamefont {Budakian}}, \bibinfo {author}
  {\bibfnamefont {S.~B.}\ \bibnamefont {Chung}}, \bibinfo {author}
  {\bibfnamefont {P.~M.}\ \bibnamefont {Goldbart}}, \ and\ \bibinfo {author}
  {\bibfnamefont {Y.}~\bibnamefont {Maeno}},\ }\href@noop {} {\bibfield
  {journal} {\bibinfo  {journal} {Science}\ }\textbf {\bibinfo {volume}
  {331}},\ \bibinfo {pages} {186} (\bibinfo {year} {2011})}\BibitemShut
  {NoStop}%
\bibitem [{\citenamefont {Fu}\ and\ \citenamefont {Kane}(2008)}]{fu2008}%
  \BibitemOpen
  \bibfield  {author} {\bibinfo {author} {\bibfnamefont {L.}~\bibnamefont
  {Fu}}\ and\ \bibinfo {author} {\bibfnamefont {C.~L.}\ \bibnamefont {Kane}},\
  }\href {\doibase 10.1103/PhysRevLett.100.096407} {\bibfield  {journal}
  {\bibinfo  {journal} {Phys. Rev. Lett.}\ }\textbf {\bibinfo {volume} {100}},\
  \bibinfo {pages} {096407} (\bibinfo {year} {2008})}\BibitemShut {NoStop}%
\bibitem [{\citenamefont {Sau}\ \emph {et~al.}(2010)\citenamefont {Sau},
  \citenamefont {Lutchyn}, \citenamefont {Tewari},\ and\ \citenamefont
  {Das~Sarma}}]{sau2010}%
  \BibitemOpen
  \bibfield  {author} {\bibinfo {author} {\bibfnamefont {J.~D.}\ \bibnamefont
  {Sau}}, \bibinfo {author} {\bibfnamefont {R.~M.}\ \bibnamefont {Lutchyn}},
  \bibinfo {author} {\bibfnamefont {S.}~\bibnamefont {Tewari}}, \ and\ \bibinfo
  {author} {\bibfnamefont {S.}~\bibnamefont {Das~Sarma}},\ }\href@noop {}
  {\bibfield  {journal} {\bibinfo  {journal} {Phys. Rev. Lett.}\ }\textbf
  {\bibinfo {volume} {104}},\ \bibinfo {pages} {040502} (\bibinfo {year}
  {2010})}\BibitemShut {NoStop}%
\bibitem [{\citenamefont {Alicea}(2010)}]{alicea2010}%
  \BibitemOpen
  \bibfield  {author} {\bibinfo {author} {\bibfnamefont {J.}~\bibnamefont
  {Alicea}},\ }\href@noop {} {\bibfield  {journal} {\bibinfo  {journal} {Phys.
  Rev. B}\ }\textbf {\bibinfo {volume} {81}},\ \bibinfo {pages} {125318}
  (\bibinfo {year} {2010})}\BibitemShut {NoStop}%
\bibitem [{\citenamefont {Lutchyn}\ \emph {et~al.}(2010)\citenamefont
  {Lutchyn}, \citenamefont {Sau},\ and\ \citenamefont
  {Das~Sarma}}]{lutchyn2010}%
  \BibitemOpen
  \bibfield  {author} {\bibinfo {author} {\bibfnamefont {R.~M.}\ \bibnamefont
  {Lutchyn}}, \bibinfo {author} {\bibfnamefont {J.~D.}\ \bibnamefont {Sau}}, \
  and\ \bibinfo {author} {\bibfnamefont {S.}~\bibnamefont {Das~Sarma}},\
  }\href@noop {} {\bibfield  {journal} {\bibinfo  {journal} {Phys. Rev. Lett.}\
  }\textbf {\bibinfo {volume} {105}},\ \bibinfo {pages} {077001} (\bibinfo
  {year} {2010})}\BibitemShut {NoStop}%
\bibitem [{\citenamefont {Oreg}\ \emph {et~al.}(2010)\citenamefont {Oreg},
  \citenamefont {Refael},\ and\ \citenamefont {von Oppen}}]{oreg2010}%
  \BibitemOpen
  \bibfield  {author} {\bibinfo {author} {\bibfnamefont {Y.}~\bibnamefont
  {Oreg}}, \bibinfo {author} {\bibfnamefont {G.}~\bibnamefont {Refael}}, \ and\
  \bibinfo {author} {\bibfnamefont {F.}~\bibnamefont {von Oppen}},\ }\href@noop
  {} {\bibfield  {journal} {\bibinfo  {journal} {Phys. Rev. Lett.}\ }\textbf
  {\bibinfo {volume} {105}},\ \bibinfo {pages} {177002} (\bibinfo {year}
  {2010})}\BibitemShut {NoStop}%
\bibitem [{\citenamefont {Mourik}\ \emph {et~al.}(2012)\citenamefont {Mourik},
  \citenamefont {Zuo}, \citenamefont {Frolov}, \citenamefont {Plissard},
  \citenamefont {Bakkers},\ and\ \citenamefont {Kouwenhoven}}]{mourik2012}%
  \BibitemOpen
  \bibfield  {author} {\bibinfo {author} {\bibfnamefont {V.}~\bibnamefont
  {Mourik}}, \bibinfo {author} {\bibfnamefont {K.}~\bibnamefont {Zuo}},
  \bibinfo {author} {\bibfnamefont {S.~M.}\ \bibnamefont {Frolov}}, \bibinfo
  {author} {\bibfnamefont {S.~R.}\ \bibnamefont {Plissard}}, \bibinfo {author}
  {\bibfnamefont {E.~P. A.~M.}\ \bibnamefont {Bakkers}}, \ and\ \bibinfo
  {author} {\bibfnamefont {L.~P.}\ \bibnamefont {Kouwenhoven}},\ }\href@noop {}
  {\bibfield  {journal} {\bibinfo  {journal} {Science}\ }\textbf {\bibinfo
  {volume} {336}},\ \bibinfo {pages} {1003} (\bibinfo {year}
  {2012})}\BibitemShut {NoStop}%
\bibitem [{\citenamefont {Kane}\ and\ \citenamefont
  {Mele}(2005{\natexlab{a}})}]{kane2005}%
  \BibitemOpen
  \bibfield  {author} {\bibinfo {author} {\bibfnamefont {C.~L.}\ \bibnamefont
  {Kane}}\ and\ \bibinfo {author} {\bibfnamefont {E.~J.}\ \bibnamefont
  {Mele}},\ }\href@noop {} {\bibfield  {journal} {\bibinfo  {journal} {Phys.
  Rev. Lett.}\ }\textbf {\bibinfo {volume} {95}} (\bibinfo {year}
  {2005}{\natexlab{a}})}\BibitemShut {NoStop}%
\bibitem [{\citenamefont {Kane}\ and\ \citenamefont
  {Mele}(2005{\natexlab{b}})}]{kane2005sf}%
  \BibitemOpen
  \bibfield  {author} {\bibinfo {author} {\bibfnamefont {C.~L.}\ \bibnamefont
  {Kane}}\ and\ \bibinfo {author} {\bibfnamefont {E.~J.}\ \bibnamefont
  {Mele}},\ }\href@noop {} {\bibfield  {journal} {\bibinfo  {journal} {Phys.
  Rev. Lett.}\ }\textbf {\bibinfo {volume} {95}},\ \bibinfo {pages} {146802}
  (\bibinfo {year} {2005}{\natexlab{b}})}\BibitemShut {NoStop}%
\bibitem [{\citenamefont {Bernevig}\ \emph {et~al.}(2006)\citenamefont
  {Bernevig}, \citenamefont {Hughes},\ and\ \citenamefont
  {Zhang}}]{bernevig2006}%
  \BibitemOpen
  \bibfield  {author} {\bibinfo {author} {\bibfnamefont {B.~A.}\ \bibnamefont
  {Bernevig}}, \bibinfo {author} {\bibfnamefont {T.~L.}\ \bibnamefont
  {Hughes}}, \ and\ \bibinfo {author} {\bibfnamefont {S.-C.}\ \bibnamefont
  {Zhang}},\ }\href@noop {} {\bibfield  {journal} {\bibinfo  {journal}
  {Science}\ }\textbf {\bibinfo {volume} {314}},\ \bibinfo {pages} {1757}
  (\bibinfo {year} {2006})}\BibitemShut {NoStop}%
\bibitem [{\citenamefont {Konig}\ \emph {et~al.}(2007)\citenamefont {Konig},
  \citenamefont {Wiedmann}, \citenamefont {Brune}, \citenamefont {Roth},
  \citenamefont {Buhmann}, \citenamefont {Molenkamp}, \citenamefont {Qi},\ and\
  \citenamefont {Zhang}}]{konig2007}%
  \BibitemOpen
  \bibfield  {author} {\bibinfo {author} {\bibfnamefont {M.}~\bibnamefont
  {Konig}}, \bibinfo {author} {\bibfnamefont {S.}~\bibnamefont {Wiedmann}},
  \bibinfo {author} {\bibfnamefont {C.}~\bibnamefont {Brune}}, \bibinfo
  {author} {\bibfnamefont {A.}~\bibnamefont {Roth}}, \bibinfo {author}
  {\bibfnamefont {H.}~\bibnamefont {Buhmann}}, \bibinfo {author} {\bibfnamefont
  {L.~W.}\ \bibnamefont {Molenkamp}}, \bibinfo {author} {\bibfnamefont {X.-L.}\
  \bibnamefont {Qi}}, \ and\ \bibinfo {author} {\bibfnamefont {S.-C.}\
  \bibnamefont {Zhang}},\ }\href {\doibase 10.1126/science.1148047} {\bibfield
  {journal} {\bibinfo  {journal} {Science}\ }\textbf {\bibinfo {volume}
  {318}},\ \bibinfo {pages} {766} (\bibinfo {year} {2007})}\BibitemShut
  {NoStop}%
\bibitem [{\citenamefont {Fu}\ \emph {et~al.}(2007)\citenamefont {Fu},
  \citenamefont {Kane},\ and\ \citenamefont {Mele}}]{fu2007prl}%
  \BibitemOpen
  \bibfield  {author} {\bibinfo {author} {\bibfnamefont {L.}~\bibnamefont
  {Fu}}, \bibinfo {author} {\bibfnamefont {C.~L.}\ \bibnamefont {Kane}}, \ and\
  \bibinfo {author} {\bibfnamefont {E.~J.}\ \bibnamefont {Mele}},\ }\href@noop
  {} {\bibfield  {journal} {\bibinfo  {journal} {Phys. Rev. Lett.}\ }\textbf
  {\bibinfo {volume} {98}},\ \bibinfo {pages} {106803} (\bibinfo {year}
  {2007})}\BibitemShut {NoStop}%
\bibitem [{\citenamefont {Moore}\ and\ \citenamefont
  {Balents}(2007)}]{moore2007}%
  \BibitemOpen
  \bibfield  {author} {\bibinfo {author} {\bibfnamefont {J.~E.}\ \bibnamefont
  {Moore}}\ and\ \bibinfo {author} {\bibfnamefont {L.}~\bibnamefont
  {Balents}},\ }\href@noop {} {\bibfield  {journal} {\bibinfo  {journal} {Phys.
  Rev. B}\ }\textbf {\bibinfo {volume} {75}},\ \bibinfo {pages} {121306}
  (\bibinfo {year} {2007})}\BibitemShut {NoStop}%
\bibitem [{\citenamefont {Roy}(2009)}]{roy2009}%
  \BibitemOpen
  \bibfield  {author} {\bibinfo {author} {\bibfnamefont {R.}~\bibnamefont
  {Roy}},\ }\href@noop {} {\bibfield  {journal} {\bibinfo  {journal} {Phys.
  Rev. B}\ }\textbf {\bibinfo {volume} {79}},\ \bibinfo {pages} {195322}
  (\bibinfo {year} {2009})}\BibitemShut {NoStop}%
\bibitem [{\citenamefont {Hasan}\ and\ \citenamefont {Kane}(2010)}]{kane2010}%
  \BibitemOpen
  \bibfield  {author} {\bibinfo {author} {\bibfnamefont {M.~Z.}\ \bibnamefont
  {Hasan}}\ and\ \bibinfo {author} {\bibfnamefont {C.~L.}\ \bibnamefont
  {Kane}},\ }\href {\doibase 10.1103/RevModPhys.82.3045} {\bibfield  {journal}
  {\bibinfo  {journal} {Rev. Mod. Phys.}\ }\textbf {\bibinfo {volume} {82}},\
  \bibinfo {pages} {3045} (\bibinfo {year} {2010})}\BibitemShut {NoStop}%
\bibitem [{\citenamefont {Hsieh}\ \emph {et~al.}(2008)\citenamefont {Hsieh},
  \citenamefont {Qian}, \citenamefont {Wray}, \citenamefont {Xia},
  \citenamefont {Hor}, \citenamefont {Cava},\ and\ \citenamefont
  {Hasan}}]{hsieh2008}%
  \BibitemOpen
  \bibfield  {author} {\bibinfo {author} {\bibfnamefont {D.}~\bibnamefont
  {Hsieh}}, \bibinfo {author} {\bibfnamefont {D.}~\bibnamefont {Qian}},
  \bibinfo {author} {\bibfnamefont {L.}~\bibnamefont {Wray}}, \bibinfo {author}
  {\bibfnamefont {Y.}~\bibnamefont {Xia}}, \bibinfo {author} {\bibfnamefont
  {Y.~S.}\ \bibnamefont {Hor}}, \bibinfo {author} {\bibfnamefont {R.~J.}\
  \bibnamefont {Cava}}, \ and\ \bibinfo {author} {\bibfnamefont {M.~Z.}\
  \bibnamefont {Hasan}},\ }\href@noop {} {\bibfield  {journal} {\bibinfo
  {journal} {Nature (London)}\ }\textbf {\bibinfo {volume} {452}},\ \bibinfo
  {pages} {970} (\bibinfo {year} {2008})}\BibitemShut {NoStop}%
\bibitem [{\citenamefont {Chen}\ \emph {et~al.}(2009)\citenamefont {Chen},
  \citenamefont {Analytis}, \citenamefont {Chu}, \citenamefont {Liu},
  \citenamefont {Mo}, \citenamefont {Qi}, \citenamefont {Zhang}, \citenamefont
  {Lu}, \citenamefont {Dai}, \citenamefont {Fang}, \citenamefont {Zhang},
  \citenamefont {Fisher}, \citenamefont {Hussain},\ and\ \citenamefont
  {Shen}}]{chengyl2009}%
  \BibitemOpen
  \bibfield  {author} {\bibinfo {author} {\bibfnamefont {Y.~L.}\ \bibnamefont
  {Chen}}, \bibinfo {author} {\bibfnamefont {J.~G.}\ \bibnamefont {Analytis}},
  \bibinfo {author} {\bibfnamefont {J.-H.}\ \bibnamefont {Chu}}, \bibinfo
  {author} {\bibfnamefont {Z.~K.}\ \bibnamefont {Liu}}, \bibinfo {author}
  {\bibfnamefont {S.-K.}\ \bibnamefont {Mo}}, \bibinfo {author} {\bibfnamefont
  {X.~L.}\ \bibnamefont {Qi}}, \bibinfo {author} {\bibfnamefont {H.~J.}\
  \bibnamefont {Zhang}}, \bibinfo {author} {\bibfnamefont {D.~H.}\ \bibnamefont
  {Lu}}, \bibinfo {author} {\bibfnamefont {X.}~\bibnamefont {Dai}}, \bibinfo
  {author} {\bibfnamefont {Z.}~\bibnamefont {Fang}}, \bibinfo {author}
  {\bibfnamefont {S.~C.}\ \bibnamefont {Zhang}}, \bibinfo {author}
  {\bibfnamefont {I.~R.}\ \bibnamefont {Fisher}}, \bibinfo {author}
  {\bibfnamefont {Z.}~\bibnamefont {Hussain}}, \ and\ \bibinfo {author}
  {\bibfnamefont {Z.-X.}\ \bibnamefont {Shen}},\ }\href@noop {} {\bibfield
  {journal} {\bibinfo  {journal} {Science}\ }\textbf {\bibinfo {volume}
  {325}},\ \bibinfo {pages} {178} (\bibinfo {year} {2009})}\BibitemShut
  {NoStop}%
\bibitem [{\citenamefont {Hsieh}\ \emph {et~al.}(2009)\citenamefont {Hsieh},
  \citenamefont {Xia}, \citenamefont {Qian}, \citenamefont {Wray},
  \citenamefont {Dil}, \citenamefont {Meier}, \citenamefont {Osterwalder},
  \citenamefont {Patthey}, \citenamefont {Checkelsky}, \citenamefont {Ong},
  \citenamefont {Lin}, \citenamefont {Bansil}, \citenamefont {Grauer},
  \citenamefont {Hor}, \citenamefont {Cava},\ and\ \citenamefont
  {Hasan}}]{hsieh2009}%
  \BibitemOpen
  \bibfield  {author} {\bibinfo {author} {\bibfnamefont {D.}~\bibnamefont
  {Hsieh}}, \bibinfo {author} {\bibfnamefont {Y.}~\bibnamefont {Xia}}, \bibinfo
  {author} {\bibfnamefont {D.}~\bibnamefont {Qian}}, \bibinfo {author}
  {\bibfnamefont {L.}~\bibnamefont {Wray}}, \bibinfo {author} {\bibfnamefont
  {J.~H.}\ \bibnamefont {Dil}}, \bibinfo {author} {\bibfnamefont
  {F.}~\bibnamefont {Meier}}, \bibinfo {author} {\bibfnamefont
  {J.}~\bibnamefont {Osterwalder}}, \bibinfo {author} {\bibfnamefont
  {L.}~\bibnamefont {Patthey}}, \bibinfo {author} {\bibfnamefont {J.~G.}\
  \bibnamefont {Checkelsky}}, \bibinfo {author} {\bibfnamefont {N.~P.}\
  \bibnamefont {Ong}}, \bibinfo {author} {\bibfnamefont {A.~V. F.~H.}\
  \bibnamefont {Lin}}, \bibinfo {author} {\bibfnamefont {A.}~\bibnamefont
  {Bansil}}, \bibinfo {author} {\bibfnamefont {D.}~\bibnamefont {Grauer}},
  \bibinfo {author} {\bibfnamefont {Y.~S.}\ \bibnamefont {Hor}}, \bibinfo
  {author} {\bibfnamefont {R.~J.}\ \bibnamefont {Cava}}, \ and\ \bibinfo
  {author} {\bibfnamefont {M.~Z.}\ \bibnamefont {Hasan}},\ }\href@noop {}
  {\bibfield  {journal} {\bibinfo  {journal} {Nature (London)}\ }\textbf
  {\bibinfo {volume} {460}},\ \bibinfo {pages} {1101} (\bibinfo {year}
  {2009})}\BibitemShut {NoStop}%
\bibitem [{\citenamefont {Volovik}(1999)}]{volovik1999}%
  \BibitemOpen
  \bibfield  {author} {\bibinfo {author} {\bibfnamefont {G.~E.}\ \bibnamefont
  {Volovik}},\ }\href@noop {} {\bibfield  {journal} {\bibinfo  {journal} {JETP
  Lett.}\ }\textbf {\bibinfo {volume} {70}},\ \bibinfo {pages} {609} (\bibinfo
  {year} {1999})}\BibitemShut {NoStop}%
\bibitem [{\citenamefont {Chiu}\ \emph {et~al.}(2011)\citenamefont {Chiu},
  \citenamefont {Gilbert},\ and\ \citenamefont {Hughes}}]{chiu2011}%
  \BibitemOpen
  \bibfield  {author} {\bibinfo {author} {\bibfnamefont {C.-K.}\ \bibnamefont
  {Chiu}}, \bibinfo {author} {\bibfnamefont {M.~J.}\ \bibnamefont {Gilbert}}, \
  and\ \bibinfo {author} {\bibfnamefont {T.~L.}\ \bibnamefont {Hughes}},\
  }\href {\doibase 10.1103/PhysRevB.84.144507} {\bibfield  {journal} {\bibinfo
  {journal} {Phys. Rev. B}\ }\textbf {\bibinfo {volume} {84}},\ \bibinfo
  {pages} {144507} (\bibinfo {year} {2011})}\BibitemShut {NoStop}%
\bibitem [{\citenamefont {Qi}\ \emph {et~al.}(2010)\citenamefont {Qi},
  \citenamefont {Hughes},\ and\ \citenamefont {Zhang}}]{qi2010}%
  \BibitemOpen
  \bibfield  {author} {\bibinfo {author} {\bibfnamefont {X.~L.}\ \bibnamefont
  {Qi}}, \bibinfo {author} {\bibfnamefont {T.~L.}\ \bibnamefont {Hughes}}, \
  and\ \bibinfo {author} {\bibfnamefont {S.~C.}\ \bibnamefont {Zhang}},\
  }\href@noop {} {\bibfield  {journal} {\bibinfo  {journal} {Phys. Rev. B}\
  }\textbf {\bibinfo {volume} {81}},\ \bibinfo {pages} {134508} (\bibinfo
  {year} {2010})}\BibitemShut {NoStop}%
\bibitem [{\citenamefont {Hor}\ \emph {et~al.}(2010)\citenamefont {Hor},
  \citenamefont {Williams}, \citenamefont {Checkelsky}, \citenamefont
  {Roushan}, \citenamefont {Seo}, \citenamefont {Xu}, \citenamefont
  {Zandbergen}, \citenamefont {Yazdani}, \citenamefont {Ong},\ and\
  \citenamefont {Cava}}]{hor2010}%
  \BibitemOpen
  \bibfield  {author} {\bibinfo {author} {\bibfnamefont {Y.~S.}\ \bibnamefont
  {Hor}}, \bibinfo {author} {\bibfnamefont {A.~J.}\ \bibnamefont {Williams}},
  \bibinfo {author} {\bibfnamefont {J.~G.}\ \bibnamefont {Checkelsky}},
  \bibinfo {author} {\bibfnamefont {P.}~\bibnamefont {Roushan}}, \bibinfo
  {author} {\bibfnamefont {J.}~\bibnamefont {Seo}}, \bibinfo {author}
  {\bibfnamefont {Q.}~\bibnamefont {Xu}}, \bibinfo {author} {\bibfnamefont
  {H.~W.}\ \bibnamefont {Zandbergen}}, \bibinfo {author} {\bibfnamefont
  {A.}~\bibnamefont {Yazdani}}, \bibinfo {author} {\bibfnamefont {N.~P.}\
  \bibnamefont {Ong}}, \ and\ \bibinfo {author} {\bibfnamefont {R.~J.}\
  \bibnamefont {Cava}},\ }\href {\doibase 10.1103/PhysRevLett.104.057001}
  {\bibfield  {journal} {\bibinfo  {journal} {Phys. Rev. Lett.}\ }\textbf
  {\bibinfo {volume} {104}},\ \bibinfo {pages} {057001} (\bibinfo {year}
  {2010})}\BibitemShut {NoStop}%
\bibitem [{\citenamefont {Hor}\ \emph {et~al.}(2011)\citenamefont {Hor},
  \citenamefont {J.G.Checkelsky}, \citenamefont {D.Qub}, \citenamefont
  {N.P.Ong},\ and\ \citenamefont {R.J.Cava}}]{hor2011}%
  \BibitemOpen
  \bibfield  {author} {\bibinfo {author} {\bibfnamefont {Y.}~\bibnamefont
  {Hor}}, \bibinfo {author} {\bibnamefont {J.G.Checkelsky}}, \bibinfo {author}
  {\bibnamefont {D.Qub}}, \bibinfo {author} {\bibnamefont {N.P.Ong}}, \ and\
  \bibinfo {author} {\bibnamefont {R.J.Cava}},\ }\href@noop {} {\bibfield
  {journal} {\bibinfo  {journal} {J. Phys. Chem. Solids.}\ }\textbf {\bibinfo
  {volume} {72}},\ \bibinfo {pages} {572} (\bibinfo {year} {2011})}\BibitemShut
  {NoStop}%
\bibitem [{\citenamefont {Wray}\ \emph {et~al.}(2011)\citenamefont {Wray},
  \citenamefont {Xu}, \citenamefont {Xia}, \citenamefont {Qian}, \citenamefont
  {Fedorov}, \citenamefont {Lin}, \citenamefont {Bansil}, \citenamefont {Fu},
  \citenamefont {Hor}, \citenamefont {Cava},\ and\ \citenamefont
  {Hasan}}]{wray2011}%
  \BibitemOpen
  \bibfield  {author} {\bibinfo {author} {\bibfnamefont {L.~A.}\ \bibnamefont
  {Wray}}, \bibinfo {author} {\bibfnamefont {S.}~\bibnamefont {Xu}}, \bibinfo
  {author} {\bibfnamefont {Y.}~\bibnamefont {Xia}}, \bibinfo {author}
  {\bibfnamefont {D.}~\bibnamefont {Qian}}, \bibinfo {author} {\bibfnamefont
  {A.~V.}\ \bibnamefont {Fedorov}}, \bibinfo {author} {\bibfnamefont
  {H.}~\bibnamefont {Lin}}, \bibinfo {author} {\bibfnamefont {A.}~\bibnamefont
  {Bansil}}, \bibinfo {author} {\bibfnamefont {L.}~\bibnamefont {Fu}}, \bibinfo
  {author} {\bibfnamefont {Y.~S.}\ \bibnamefont {Hor}}, \bibinfo {author}
  {\bibfnamefont {R.~J.}\ \bibnamefont {Cava}}, \ and\ \bibinfo {author}
  {\bibfnamefont {M.~Z.}\ \bibnamefont {Hasan}},\ }\href {\doibase
  10.1103/PhysRevB.83.224516} {\bibfield  {journal} {\bibinfo  {journal} {Phys.
  Rev. B}\ }\textbf {\bibinfo {volume} {83}},\ \bibinfo {pages} {224516}
  (\bibinfo {year} {2011})}\BibitemShut {NoStop}%
\bibitem [{\citenamefont {Fu}\ and\ \citenamefont {Berg}(2010)}]{fu2010}%
  \BibitemOpen
  \bibfield  {author} {\bibinfo {author} {\bibfnamefont {L.}~\bibnamefont
  {Fu}}\ and\ \bibinfo {author} {\bibfnamefont {E.}~\bibnamefont {Berg}},\
  }\href {\doibase 10.1103/PhysRevLett.105.097001} {\bibfield  {journal}
  {\bibinfo  {journal} {Phys. Rev. Lett.}\ }\textbf {\bibinfo {volume} {105}},\
  \bibinfo {pages} {097001} (\bibinfo {year} {2010})}\BibitemShut {NoStop}%
\bibitem [{\citenamefont {Zhang}\ \emph {et~al.}(2009)\citenamefont {Zhang},
  \citenamefont {Liu}, \citenamefont {Qi}, \citenamefont {Dai}, \citenamefont
  {Fang},\ and\ \citenamefont {Zhang}}]{zhang2009}%
  \BibitemOpen
  \bibfield  {author} {\bibinfo {author} {\bibfnamefont {H.}~\bibnamefont
  {Zhang}}, \bibinfo {author} {\bibfnamefont {C.-X.}\ \bibnamefont {Liu}},
  \bibinfo {author} {\bibfnamefont {X.-L.}\ \bibnamefont {Qi}}, \bibinfo
  {author} {\bibfnamefont {X.}~\bibnamefont {Dai}}, \bibinfo {author}
  {\bibfnamefont {Z.}~\bibnamefont {Fang}}, \ and\ \bibinfo {author}
  {\bibfnamefont {S.-C.}\ \bibnamefont {Zhang}},\ }\href@noop {} {\bibfield
  {journal} {\bibinfo  {journal} {Nat. Phys.}\ }\textbf {\bibinfo {volume}
  {5}},\ \bibinfo {pages} {438} (\bibinfo {year} {2009})}\BibitemShut {NoStop}%
\bibitem [{\citenamefont {Liu}\ \emph {et~al.}(2010)\citenamefont {Liu},
  \citenamefont {Qi}, \citenamefont {Zhang}, \citenamefont {Dai}, \citenamefont
  {Fang},\ and\ \citenamefont {Zhang}}]{liu2010}%
  \BibitemOpen
  \bibfield  {author} {\bibinfo {author} {\bibfnamefont {C.-X.}\ \bibnamefont
  {Liu}}, \bibinfo {author} {\bibfnamefont {X.-L.}\ \bibnamefont {Qi}},
  \bibinfo {author} {\bibfnamefont {H.}~\bibnamefont {Zhang}}, \bibinfo
  {author} {\bibfnamefont {X.}~\bibnamefont {Dai}}, \bibinfo {author}
  {\bibfnamefont {Z.}~\bibnamefont {Fang}}, \ and\ \bibinfo {author}
  {\bibfnamefont {S.-C.}\ \bibnamefont {Zhang}},\ }\href {\doibase
  10.1103/PhysRevB.82.045122} {\bibfield  {journal} {\bibinfo  {journal} {Phys.
  Rev. B}\ }\textbf {\bibinfo {volume} {82}},\ \bibinfo {pages} {045122}
  (\bibinfo {year} {2010})}\BibitemShut {NoStop}%
\bibitem [{\citenamefont {Gennes}(1966)}]{degennes1966}%
  \BibitemOpen
  \bibfield  {author} {\bibinfo {author} {\bibfnamefont {P.~D.}\ \bibnamefont
  {Gennes}},\ }\href@noop {} {\emph {\bibinfo {title} {Superconductivity of
  metals and alloys}}}\ (\bibinfo  {publisher} {W. A. Benjamin Inc.},\ \bibinfo
  {year} {1966})\BibitemShut {NoStop}%
\bibitem [{\citenamefont {Vafek}\ \emph {et~al.}(2001)\citenamefont {Vafek},
  \citenamefont {Melikyan}, \citenamefont {Franz},\ and\ \citenamefont
  {Te\ifmmode \check{s}\else \v{s}\fi{}anovi\ifmmode~\acute{c}\else
  \'{c}\fi{}}}]{vafek2001}%
  \BibitemOpen
  \bibfield  {author} {\bibinfo {author} {\bibfnamefont {O.}~\bibnamefont
  {Vafek}}, \bibinfo {author} {\bibfnamefont {A.}~\bibnamefont {Melikyan}},
  \bibinfo {author} {\bibfnamefont {M.}~\bibnamefont {Franz}}, \ and\ \bibinfo
  {author} {\bibfnamefont {Z.}~\bibnamefont {Te\ifmmode \check{s}\else
  \v{s}\fi{}anovi\ifmmode~\acute{c}\else \'{c}\fi{}}},\ }\href {\doibase
  10.1103/PhysRevB.63.134509} {\bibfield  {journal} {\bibinfo  {journal} {Phys.
  Rev. B}\ }\textbf {\bibinfo {volume} {63}},\ \bibinfo {pages} {134509}
  (\bibinfo {year} {2001})}\BibitemShut {NoStop}%
\bibitem [{\citenamefont {Wang}\ and\ \citenamefont
  {MacDonald}(1995)}]{wang1995}%
  \BibitemOpen
  \bibfield  {author} {\bibinfo {author} {\bibfnamefont {Y.}~\bibnamefont
  {Wang}}\ and\ \bibinfo {author} {\bibfnamefont {A.~H.}\ \bibnamefont
  {MacDonald}},\ }\href {\doibase 10.1103/PhysRevB.52.R3876} {\bibfield
  {journal} {\bibinfo  {journal} {Phys. Rev. B}\ }\textbf {\bibinfo {volume}
  {52}},\ \bibinfo {pages} {R3876} (\bibinfo {year} {1995})}\BibitemShut
  {NoStop}%
\bibitem [{\citenamefont {Zak}(1964{\natexlab{a}})}]{zak1964}%
  \BibitemOpen
  \bibfield  {author} {\bibinfo {author} {\bibfnamefont {J.}~\bibnamefont
  {Zak}},\ }\href {\doibase 10.1103/PhysRev.134.A1602} {\bibfield  {journal}
  {\bibinfo  {journal} {Phys. Rev.}\ }\textbf {\bibinfo {volume} {134}},\
  \bibinfo {pages} {A1602} (\bibinfo {year} {1964}{\natexlab{a}})}\BibitemShut
  {NoStop}%
\bibitem [{\citenamefont {Zak}(1964{\natexlab{b}})}]{zak1964b}%
  \BibitemOpen
  \bibfield  {author} {\bibinfo {author} {\bibfnamefont {J.}~\bibnamefont
  {Zak}},\ }\href {\doibase 10.1103/PhysRev.134.A1607} {\bibfield  {journal}
  {\bibinfo  {journal} {Phys. Rev.}\ }\textbf {\bibinfo {volume} {134}},\
  \bibinfo {pages} {A1607} (\bibinfo {year} {1964}{\natexlab{b}})}\BibitemShut
  {NoStop}%
\bibitem [{\citenamefont {Han}(2010)}]{han2010}%
  \BibitemOpen
  \bibfield  {author} {\bibinfo {author} {\bibfnamefont {Q.}~\bibnamefont
  {Han}},\ }\href@noop {} {\bibfield  {journal} {\bibinfo  {journal} {J. Phys.:
  Condens. Matter}\ }\textbf {\bibinfo {volume} {22}},\ \bibinfo {pages}
  {035702} (\bibinfo {year} {2010})}\BibitemShut {NoStop}%
\bibitem [{\citenamefont {Hung}\ \emph {et~al.}(2012)\citenamefont {Hung},
  \citenamefont {Song}, \citenamefont {Chen}, \citenamefont {Ma}, \citenamefont
  {Xue},\ and\ \citenamefont {Wu}}]{hung2012}%
  \BibitemOpen
  \bibfield  {author} {\bibinfo {author} {\bibfnamefont {H.-H.}\ \bibnamefont
  {Hung}}, \bibinfo {author} {\bibfnamefont {C.-L.}\ \bibnamefont {Song}},
  \bibinfo {author} {\bibfnamefont {X.}~\bibnamefont {Chen}}, \bibinfo {author}
  {\bibfnamefont {X.}~\bibnamefont {Ma}}, \bibinfo {author} {\bibfnamefont
  {Q.-k.}\ \bibnamefont {Xue}}, \ and\ \bibinfo {author} {\bibfnamefont
  {C.}~\bibnamefont {Wu}},\ }\href {\doibase 10.1103/PhysRevB.85.104510}
  {\bibfield  {journal} {\bibinfo  {journal} {Phys. Rev. B}\ }\textbf {\bibinfo
  {volume} {85}},\ \bibinfo {pages} {104510} (\bibinfo {year}
  {2012})}\BibitemShut {NoStop}%
\bibitem [{\citenamefont {B\'eri}\ and\ \citenamefont
  {Cooper}(2011)}]{beri2011}%
  \BibitemOpen
  \bibfield  {author} {\bibinfo {author} {\bibfnamefont {B.}~\bibnamefont
  {B\'eri}}\ and\ \bibinfo {author} {\bibfnamefont {N.~R.}\ \bibnamefont
  {Cooper}},\ }\href@noop {} {\bibfield  {journal} {\bibinfo  {journal} {Phys.
  Rev. Lett.}\ }\textbf {\bibinfo {volume} {107}},\ \bibinfo {pages} {145301}
  (\bibinfo {year} {2011})}\BibitemShut {NoStop}%
\bibitem [{\citenamefont {Yasui}\ and\ \citenamefont {Kita}(1999)}]{yasui1999}%
  \BibitemOpen
  \bibfield  {author} {\bibinfo {author} {\bibfnamefont {K.}~\bibnamefont
  {Yasui}}\ and\ \bibinfo {author} {\bibfnamefont {T.}~\bibnamefont {Kita}},\
  }\href@noop {} {\bibfield  {journal} {\bibinfo  {journal} {Phys. Rev. Lett.}\
  }\textbf {\bibinfo {volume} {83}},\ \bibinfo {pages} {4168} (\bibinfo {year}
  {1999})}\BibitemShut {NoStop}%
\bibitem [{\citenamefont {Shoemake}\ \emph {et~al.}(1969)\citenamefont
  {Shoemake}, \citenamefont {Rayne},\ and\ \citenamefont {Ure}}]{shoemake1969}%
  \BibitemOpen
  \bibfield  {author} {\bibinfo {author} {\bibfnamefont {G.~E.}\ \bibnamefont
  {Shoemake}}, \bibinfo {author} {\bibfnamefont {J.~A.}\ \bibnamefont {Rayne}},
  \ and\ \bibinfo {author} {\bibfnamefont {R.~W.}\ \bibnamefont {Ure}},\
  }\href@noop {} {\bibfield  {journal} {\bibinfo  {journal} {Phys. Rev.}\
  }\textbf {\bibinfo {volume} {185}},\ \bibinfo {pages} {1046} (\bibinfo {year}
  {1969})}\BibitemShut {NoStop}%
\bibitem [{\citenamefont {Wray}\ \emph {et~al.}(2010)\citenamefont {Wray},
  \citenamefont {Xu}, \citenamefont {Xia}, \citenamefont {Hor}, \citenamefont
  {Qian}, \citenamefont {Fedorov}, \citenamefont {Lin}, \citenamefont {Bansil},
  \citenamefont {Cava},\ and\ \citenamefont {Hasan}}]{wray2010}%
  \BibitemOpen
  \bibfield  {author} {\bibinfo {author} {\bibfnamefont {L.~A.}\ \bibnamefont
  {Wray}}, \bibinfo {author} {\bibfnamefont {S.-Y.}\ \bibnamefont {Xu}},
  \bibinfo {author} {\bibfnamefont {Y.}~\bibnamefont {Xia}}, \bibinfo {author}
  {\bibfnamefont {Y.~S.}\ \bibnamefont {Hor}}, \bibinfo {author} {\bibfnamefont
  {D.}~\bibnamefont {Qian}}, \bibinfo {author} {\bibfnamefont {A.~V.}\
  \bibnamefont {Fedorov}}, \bibinfo {author} {\bibfnamefont {H.}~\bibnamefont
  {Lin}}, \bibinfo {author} {\bibfnamefont {A.}~\bibnamefont {Bansil}},
  \bibinfo {author} {\bibfnamefont {R.~J.}\ \bibnamefont {Cava}}, \ and\
  \bibinfo {author} {\bibfnamefont {M.~Z.}\ \bibnamefont {Hasan}},\ }\href@noop
  {} {\bibfield  {journal} {\bibinfo  {journal} {Nat. Phys.}\ }\textbf
  {\bibinfo {volume} {6}},\ \bibinfo {pages} {855} (\bibinfo {year}
  {2010})}\BibitemShut {NoStop}%
\bibitem [{\citenamefont {Chiu}\ \emph {et~al.}(2012)\citenamefont {Chiu},
  \citenamefont {Ghaemi},\ and\ \citenamefont {Hughes}}]{chiughaemihughes}%
  \BibitemOpen
  \bibfield  {author} {\bibinfo {author} {\bibfnamefont {C.-K.}\ \bibnamefont
  {Chiu}}, \bibinfo {author} {\bibfnamefont {P.}~\bibnamefont {Ghaemi}}, \ and\
  \bibinfo {author} {\bibfnamefont {T.~L.}\ \bibnamefont {Hughes}},\
  }\href@noop {} {\bibfield  {journal} {\bibinfo  {journal}
  {{a}rXiv:1203.2958}\ } (\bibinfo {year} {2012})}\BibitemShut {NoStop}%
\end{thebibliography}%


















\end{document}